\documentclass[prl,twocolumn,superscriptaddress,titleinbib,notitlepage]{revtex4-2}

\usepackage{etoolbox}
\newbool{isFinal}

\boolfalse{isFinal}

\pdfoutput=1
\usepackage{makeidx}
\usepackage[utf8]{inputenc}
\usepackage{graphicx}
\usepackage{amsmath}
\usepackage{amssymb}
\usepackage{mathrsfs}
\usepackage{subfigure}
\usepackage{dcolumn}
\usepackage{bm}
\usepackage{bbm}
\usepackage{color}
\usepackage{esint}
\usepackage{lineno}
\usepackage[breaklinks,
            colorlinks,
            urlcolor=blue,
            linkcolor=blue,
            anchorcolor=blue,
            citecolor=blue]{hyperref}

\usepackage{xcolor}
\definecolor{jfColor}{HTML}{FF00FF}

\definecolor{jbColor}{HTML}{33AAFF}

\usepackage{ulem}

\usepackage{orcidlink}

\ifbool{isFinal}{
    \usepackage[final]{changes}

}{
    \usepackage[markup=underlined]{changes}

}

\begin{document}

\title{Generation and optimization of entanglement \\ between atoms chirally coupled to spin cavities}

\author{Jia-Bin You\orcidlink{0000-0001-8815-1855}}
\email{you\_jiabin@ihpc.a-star.edu.sg}
\affiliation{Institute of High Performance Computing (IHPC), Agency for Science, Technology and Research (A*STAR), 1 Fusionopolis Way, \#16-16 Connexis, Singapore 138632, Republic of Singapore}

\author{Jian Feng Kong\orcidlink{0000-0001-5980-4140}}
\affiliation{Institute of High Performance Computing (IHPC), Agency for Science, Technology and Research (A*STAR), 1 Fusionopolis Way, \#16-16 Connexis, Singapore 138632, Republic of Singapore}

\author{Davit Aghamalyan}
\affiliation{Institute of High Performance Computing (IHPC), Agency for Science, Technology and Research (A*STAR), 1 Fusionopolis Way, \#16-16 Connexis, Singapore 138632, Republic of Singapore}

\author{Wai-Keong Mok}
\affiliation{Institute for Quantum Information and Matter, California Institute of Technology, Pasadena, CA 91125, USA}

\author{Kian Hwee Lim}
\affiliation{Centre for Quantum Technologies, National University of Singapore, 3 Science Drive 2, Singapore 117543}

\author{Jun Ye\orcidlink{0000-0003-1963-0865}}
\affiliation{Institute of High Performance Computing (IHPC), Agency for Science, Technology and Research (A*STAR), 1 Fusionopolis Way, \#16-16 Connexis, Singapore 138632, Republic of Singapore}

\author{Ching Eng Png}
\affiliation{Institute of High Performance Computing (IHPC), Agency for Science, Technology and Research (A*STAR), 1 Fusionopolis Way, \#16-16 Connexis, Singapore 138632, Republic of Singapore}

\author{Francisco J. Garc\'{i}a-Vidal\orcidlink{0000-0003-4354-0982}}
\email{fj.garcia@uam.es}
\affiliation{Institute of High Performance Computing (IHPC), Agency for Science, Technology and Research (A*STAR), 1 Fusionopolis Way, \#16-16 Connexis, Singapore 138632, Republic of Singapore}
\affiliation{Departamento de F\'{i}sica Te\'{o}rica de la Materia Condensada and Condensed Matter Physics Center (IFIMAC),
Universidad Aut\'{o}noma de Madrid, E-28049 Madrid, Spain}

\begin{abstract}

We explore the generation and optimization of entanglement between atoms chirally coupled to finite 1D spin chains, functioning as {\it spin cavities}.
By diagonalizing the spin cavity Hamiltonian, we identify a parity effect that influences entanglement, with small even-sized cavities chirally coupled to atoms expediting entanglement generation by approximately $50\%$ faster than non-chiral coupling.
Applying a classical driving field to the atoms reveals oscillations in concurrence, with resonant dips at specific driving strengths due to the resonances between the driven atom and the spin cavity.
Extending our study to systems with energetic disorder, we find that high concurrence can be achieved regardless of disorder strength when the inverse participation ratio of the resulting eigenstates is favorable.
Finally, we demonstrate that controlled disorder within the cavity significantly enhances and expedites entanglement generation, achieving higher concurrences up to four times faster than those attained in ordered systems.

\end{abstract}

\maketitle

\textit{Introduction.---} Entanglement is a vital resource for various quantum information technologies, such as quantum teleportation \cite{PhysRevLett.70.1895}, quantum key distribution \cite{PhysRevLett.67.661,Nature.421.238}, and quantum computing \cite{Nature.604.457}.
Researchers have been actively working to develop techniques for generating and optimizing entanglement in a variety of physical systems, such as plasmonic structures \cite{PhysRevLett.106.020501}, cavity quantum electrodynamics (QED) setups \cite{RevModPhys.73.565}, and circuit-QED
platforms \cite{RevModPhys.93.025005}.
One of the key figures of merit for entanglement generation is to maximize the entanglement while minimizing the generation time \cite{goerz2015optimizing,watts2015optimizing,muller2011optimizing}. In recent years, chiral light-matter interaction has turned out to be a promising ingredient   \cite{pichler2015quantum,PhysRevA.93.062104,ramos2014quantum,lodahl2017chiral} for generating non-classical states of light \cite{kleinbeck2023creation} and entangled states, and several theoretical studies have elucidated that fully chiral systems can significantly boost dynamical entanglement generation \cite{gonzalez2015chiral,PhysRevA.101.053861} and both quantum state
\cite{PhysRevA.101.053861,PhysRevResearch.2.013369} and entanglement transfer \cite{PhysRevResearch.2.013369}.
Chirality appears as a natural manifestation of spin-orbit coupling of light \cite{PhysRevA.93.062104}, as demonstrated in seminal experiments with atoms and quantum dots coupled to photonic nanostructures \cite{doi:10.1126/science.1257671,PhysRevLett.115.153901,doi:10.1126/science.1233746,PhysRevA.84.042341,lodahl2017chiral,yang2016long}.

In this Letter, we theoretically explore the generation and optimization of entanglement between atoms chirally coupled to finite 1D spin chains that act as {\it spin cavities}. In the search for new platforms to
achieve robust entangled states,  our idea is replacing the photons in standard optical cavities by the magnonic excitations supported by spin chains. By employing a variational matrix product state (MPS) algorithm,
which enables us to investigate phenomena beyond mean-field theory and surpasses the limitations of the Born-Markov approximation,
we create a complete map of entanglement generation in these platforms, using the concurrence as a measure of the degree of entanglement.
We first study the influence of entanglement on cavity length and chirality, showing that parity is critical, largely favouring even-sized cavities versus odd ones to
reach concurrences close to $1$, i.e., maximum entanglement. When applying a classical driving field to the system, we observe that, contrary to what is expected, the concurrence does not decrease monotonically with the increase of the driving field but shows
an oscillatory behaviour.  We also extend our study to more practical situations by considering how disorder affects entanglement in these chiral systems. Surprisingly, we observe that high values of concurrence can be attained even in conditions of strong disorder, with the inverse participation ratio of the resulting eigenstates being the key factor that determines the degree of entanglement. Enlightened by this finding, we finally consider leveraging ``disorder'' in the cavity to both maximize the degree of entanglement and expedite the generation time, showing that a higher concurrence can be achieved within a shorter period of time when on-site energies and hopping parameters in the spin cavity are properly tuned.

\textit{Model.---} We consider entanglement generation between atoms coupled to a spin cavity, as shown in Fig. \ref{model}(a). A series of $N$ two-level systems describe the atoms, with interactions mediated by magnons excited within the spin cavity, consisting of $L$ spin sites. The spin cavity Hamiltonian is $H_{c}=\sum_{i=1}^{L}\Delta_{c_{i}}\sigma_{c_{i}}^{+}\sigma_{c_{i}}^{-}+\sum_{i=1}^{L-1}J_{c_{i}}(\sigma_{c_{i+1}}^{+}\sigma_{c_{i}}^{-}+\sigma_{c_{i}}^{+}\sigma_{c_{i+1}}^{-})$. The atomic Hamiltonian is $H_{n}=\sum_{i=1}^{N}\Delta_{n_{i}}\sigma_{n_{i}}^{+}\sigma_{n_{i}}^{-}+\Omega_{n_{i}}(\sigma_{n_{i}}^{+}+\sigma_{n_{i}}^{-})$, where $\Delta_{c_{i}(n_{i})}$ represents the on-site energy, $J_{c_{i}}$ is the hopping strength for the spin cavity, $\Omega_{n_{i}}$ is the driving strength for the atoms, and $\sigma_{c_{i}(n_{i})}^{\pm}$ are the raising and lowering operators of the spin cavity and atoms. The chiral coupling between atoms and the spin cavity is $H_{g}=\sum_{i=1}^{N}g_{n_{i}}e^{-i\phi_{n_{i}}}(\sigma_{n_{i}}^{+}\sigma_{c_{\mathcal{L}[n_{i}]}}^{-}+\sigma_{c_{\mathcal{R}[n_{i}]}}^{+}\sigma_{n_{i}}^{-})+\text{h.c.}$, where $g_{n_{i}}$ and $\phi_{n_{i}}$ are the coupling strength and hopping phase, and $\mathcal{L}[n_{i}]$ and $\mathcal{R}[n_{i}]$ represent the number of cavity spins to the left and right of atom $n_{i}$. These can be used to label the positions of the atoms along the cavity. Chirality is introduced through the hopping phase $\phi_{n_{i}}$, related to the Lamb shift between neighboring coupling points for atom $n_{i}$ \cite{PhysRevA.90.013837}. For simplicity, we analyze the case of just two atoms at positions $n_1$ and $n_2$, resonant with the spin cavity, with $\Delta_{c_{i}}=\Delta_{n_{i}}=\Delta=0$ and uniform hopping strength, $J_{c_{i}}=J$. The chiral couplings are uniform, $g_{n_{i}}=g$ and $\phi_{n_{i}}=\phi$. By the unitary transformation $\sigma_{c_{i}}^{-}=\sqrt{\frac{2}{L+1}}\sum_{k=1}^{L}\sin{\frac{\pi k i}{L+1}}\eta_{c,k}$, the spin cavity Hamiltonian is diagonalized as $H_{c}=\sum_{k=1}^{L}\epsilon_{c,k}\eta_{c,k}^{\dag}\eta_{c,k}$, where $\eta_{c,k}$ is the magnonic mode in the spin cavity, and its dispersion is $\epsilon_{c,k}=\Delta+2J\cos\frac{\pi k}{L+1}$. More details on this model Hamiltonian, the dispersion relation of the magnons, and the relationship between chirality and the hopping phase are in Section I of the SM \cite{sm}. As commented above, we employ a variational MPS algorithm \cite{SCHOLLWOCK201196,Verstraete2008,PhysRevLett.114.220601,PhysRevA.103.053517,PhysRevLett.107.070601,PhysRevB.94.165116} to exactly account for the time evolution of density matrix of the whole system. Dissipation could also be incorporated into the numerical framework and a brief discussion on the effects of internal losses can be found in Section II of the SM \cite{sm}.

\begin{figure}[htbp]
\includegraphics[width=9cm]{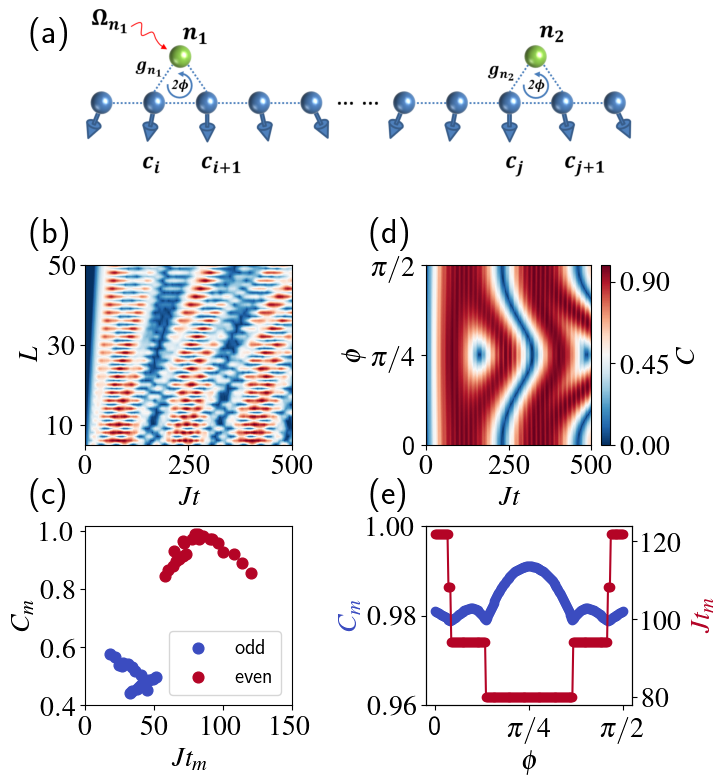}\\
\caption{(a) Schematic of driven atoms coupled to a spin cavity. Panel (b) renders concurrence versus time and $L$ (from $5$ to $50$), whereas panel (c) shows the maximum concurrence, $C_{m}$, and the corresponding time $t_{m}$ (in units of $1/J$) for the cases in (b). Here, $\mathcal{L}[n_{1}]=2$ and $\mathcal{L}[n_{2}]=L-2$, with hopping phases for atoms $n_1$ and $n_2$ set to $\phi=\pi/4$. Panels (d) and (e) illustrate chirality effect with cavity length fixed to $L=6$: (d) concurrence versus hopping phase $\phi$ and time, and (e) $C_{m}$ (blue line) and $Jt_{m}$ (red line) versus $\phi$. Panels (b) and (d) share the same colorbar. The amplitude is $g/J=0.1$. Bond dimension $D=10$ is used for MPS simulations.}
\label{model}
\end{figure}

\textit{Parity and chirality.---} We first investigate the parity effects of cavity length and atom positions, and chirality of the hopping phase on the concurrence of the two atoms. Concurrence ($C$) is a measure of the degree of entanglement \cite{wootters2001entanglement,RevModPhys.81.865}, varying between $C=1$ (maximum entanglement) and $C=0$ (no entanglement).
By performing a partial trace over the cavity spins for the density matrix $\rho$ of the whole system, we obtain the reduced density matrix $\rho_{n}=\text{tr}_{c}(\rho)$ describing a mixed state of the two atoms $n_{1}$ and $n_{2}$.
The concurrence of the reduced density matrix $\rho_{n}$ is defined as $C=\max\{0,\sqrt{\lambda_{1}}-\sqrt{\lambda_{2}}-\sqrt{\lambda_{3}}-\sqrt{\lambda_{4}}\}$, where $\lambda_{1},\cdots,\lambda_{4}$ are the eigenvalues, in decreasing order,
of the matrix $\sqrt{ \sqrt{\rho_n} \tilde{\rho}_n \sqrt{\rho_n}}$. Here, $\tilde{\rho}_{n}=(\sigma_{y}\otimes\sigma_{y})\rho_{n}^{*}(\sigma_{y}\otimes\sigma_{y})$ is the spin-flipped state of $\rho_{n}$, and $\sigma_{y}$ is the Pauli matrix.

For the numerical simulations we present next, the atoms are initially prepared with atom $n_{1}$ in the excited state $|1\rangle$ and atom $n_{2}$ in the ground state $|0\rangle$, with no external driving ($\Omega_{n_{i}}=0$),
whereas all the cavity spins are in the ground state. In this scenario, the system can be restricted to the single excitation subspace. We plot the concurrence versus time, $Jt$, and cavity length, $L$, in Fig. \ref{model}(b) for a particular chirality ($\phi=\pi/4$), observing a zero-concurrence region in the leftmost part of the panel that indicates ballistic propagation of excitation at short times.
Rabi oscillations occur during the time evolution of concurrence due to the reflection of excitation within the spin cavity. Importantly, concurrence oscillates along the $L$-axis following the parity of the cavity length.  This trend is clearly seen in Fig. \ref{model}(c), which shows the maximum concurrence, $C_{m}$ at the first occurrence and the corresponding optimal time, $Jt_{m}$, for different cavity lengths.
We find that cavities with an odd number of spins result in low concurrence ($C_{m}<0.6$), while those with an even number of spins lead to high concurrence ($C_{m}>0.8$).
The maximum $C_{m}$ of approximately $0.99$ occurs for $L=10$. Detailed information on the cavity length associated with each point in Fig. \ref{model}(c) can be found in Section I of the SM \cite{sm}.
Furthermore, for cavities with the same parity, Fig. \ref{model}(b) shows that $C_{m}$ tends to decrease as the cavity length increases.
We can understand this behaviour observed in the numerical results by analyzing the dispersion relation of the magnons excited in the spin cavity, $\epsilon_{c,k}$.
First, we observe that for an odd-sized cavity, the eigenmode with $k=(L+1)/2$ has an energy of $\Delta$, which is resonant with the atoms. In contrast, for an even-sized cavity, there is no cavity mode resonant with the atoms. The leakage of excitation from the atoms to the spin cavity via the resonant magnonic mode then leads to low concurrence in odd-sized cavities. Second, because of the increasing number of cavity modes that become near-resonant with the atomic modes as cavity's length is increased, greater leakage of the atom excitations into the spin cavity appears for larger cavities, reducing the entanglement between the two atoms.

In the search for physical insight into this problem, given that the coupling strengths between the spin cavity and the atoms are much smaller than the cavity's energy scale ($g \ll J$), perturbation theory can be applied to derive an effective Hamiltonian of the atomic subsystem for both even and odd-sized cavities.  A detailed account of this approach based on perturbation theory and its comparison with the numerical results can be found in Section III of the SM \cite{sm}.
For even-sized cavities, we can obtain analytical results. When the distance between the atoms, $\Delta{n}=|\mathcal{L}[n_1] - \mathcal{L}[n_2]|$, is even, the concurrence is $C(t)=|\sin{\frac{2g^{2}}{J}t}|$. When the distance is odd, the concurrence is $C(t)=2\sqrt{[1-\alpha(t)]\alpha(t)}$, where $\alpha(t)=\sin^{2}{\frac{g^{2}\sqrt{1+\cos^{2}{2\phi}}t}{J}}/(1+\cos^{2}{2\phi})$, which depends on the chirality through the hopping phase $\phi$.
We observe that for even-sized cavities, the maximum concurrence can reach $1$ in both cases, as our MPS numerical results show.

We then study the effect of chirality by analyzing the dependence of concurrence on the hopping phase $\phi$.
According to the perturbation theory discussed above, the concurrence between two atoms with an even distance is independent of the phase $\phi$, while it does depend on $\phi$ when the distance is odd.
To highlight the advantage of chirality, we consider a small, even-sized cavity with $L = 6$ coupled to two atoms with an odd distance, $\mathcal{L}[n_{1}] = 2$ and $\mathcal{L}[n_{2}] = 5$. The dependence of the concurrence versus
$\phi$ and time is depicted in Fig. \ref{model}(d).
First, as expected, the pattern is symmetric with respect to $\phi=\pi/4$. As the phase $\phi$ changes from $0$ to $\pi/4$, the time required for the first occurrence of maximum concurrence becomes shorter.
Fig. \ref{model}(e) shows that chiral coupling results in higher entanglement within a shorter time, achieving this approximately $50\%$ faster than non-chiral coupling ($\phi=0$).
Based on our numerical results and the insight provided by our analytical approach, we can conclude that short,
even-sized cavities displaying a $\phi=\pi/4$ chiral atom-cavity coupling are the optimal configurations to achieve the higher degree of entanglement between the two atoms.

\begin{figure}[htp]
\includegraphics[width=8.5cm]{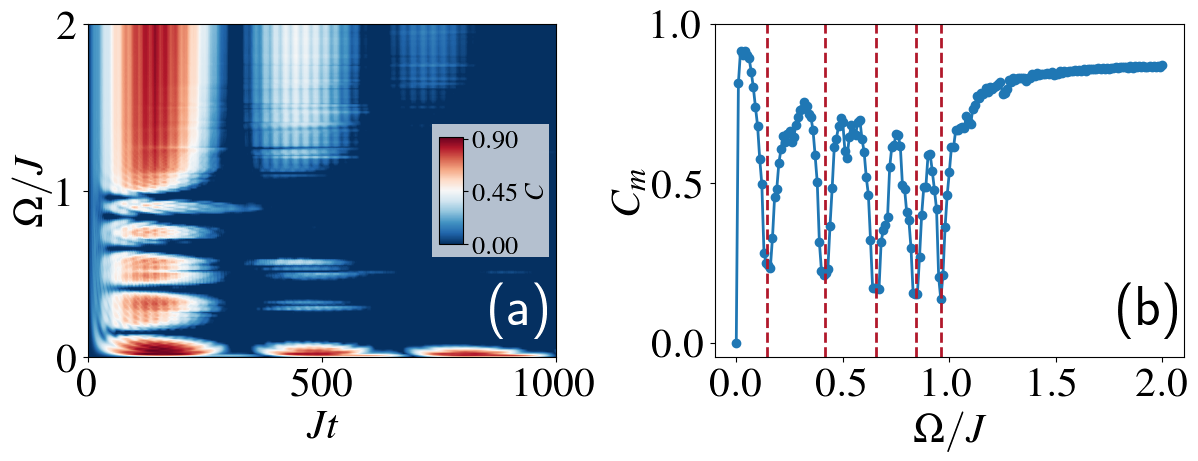}\\
\caption{Effect of driving on concurrence. Parameters are $L=10$, $\mathcal{L}[n_{1}]=2$, $\mathcal{L}[n_{2}]=8$, $g/J=0.1$, and $\phi=\pi/4$. The on-site energies and hopping amplitudes are the same as in Fig. \ref{model}. (a) Concurrence versus time $Jt$ and driving $\Omega/J$; (b) maximum concurrence versus driving $\Omega/J$.}
\label{driving}
\end{figure}

\textit{Effect of driving.---} In light of the previous findings, we focus now on an even-sized spin cavity with number $L=10$, chirally-coupled ($\phi=\pi/4$) to two atoms.
We assume that the atom $n_1$ is driven by a classical driving field with strength $\Omega_{n_{1}}\equiv\Omega$.
Here, as a difference with the case analyzed in Fig. \ref{model}, we consider that both atoms are in their ground states at $t=0$.
Our results in Fig. \ref{driving}(a) reveal that entanglement dynamically evolves showing revival-and-death phenomena \cite{doi:10.1126/science.1167343}, with decreasing concurrence in each oscillatory lobe.
When looking at the maximum concurrence $C_{m}$ versus driving strength $\Omega$ (Fig. \ref{driving}(b)), we find that entanglement exhibits an oscillatory behavior too. Specifically, there are dips in $C_{m}$ at certain driving strengths $\Omega/J<1$.
As previously mentioned, for even-sized cavities, there is no cavity mode resonant with the undriven atoms, resulting in high concurrences.
However, by driving atom $n_{1}$, it is possible to tune its excitation energy to be resonant with the cavity modes, leading to the observed concurrence dips.
The driving strength for this resonance condition is $\Omega=|J\cos\frac{\pi k}{L+1}|$.
The positions of these driving strengths (red-dashed lines) correspond exactly to the locations of the dips observed in the numerical simulations shown in Fig. \ref{driving}(b).
More details on the derivation for this analytical formula for the maximum concurrence dips can be found in Section IV of the SM \cite{sm}.
Surprisingly, the highest entanglement scenario occurs at very weak driving strength ($C_{m}\approx0.92$ at $\Omega/J=0.02$). As the driving strength increases beyond $J$, the maximum concurrence gradually saturates to a high value, which is nevertheless lower than that obtained for weak driving. Our findings provide deep insights into the intricate and non-trivial interplay between driving and entanglement generation, showing that a weak external driving leads to a higher entanglement.

\begin{figure}[htp]
\centering
\includegraphics[width=8.5cm]{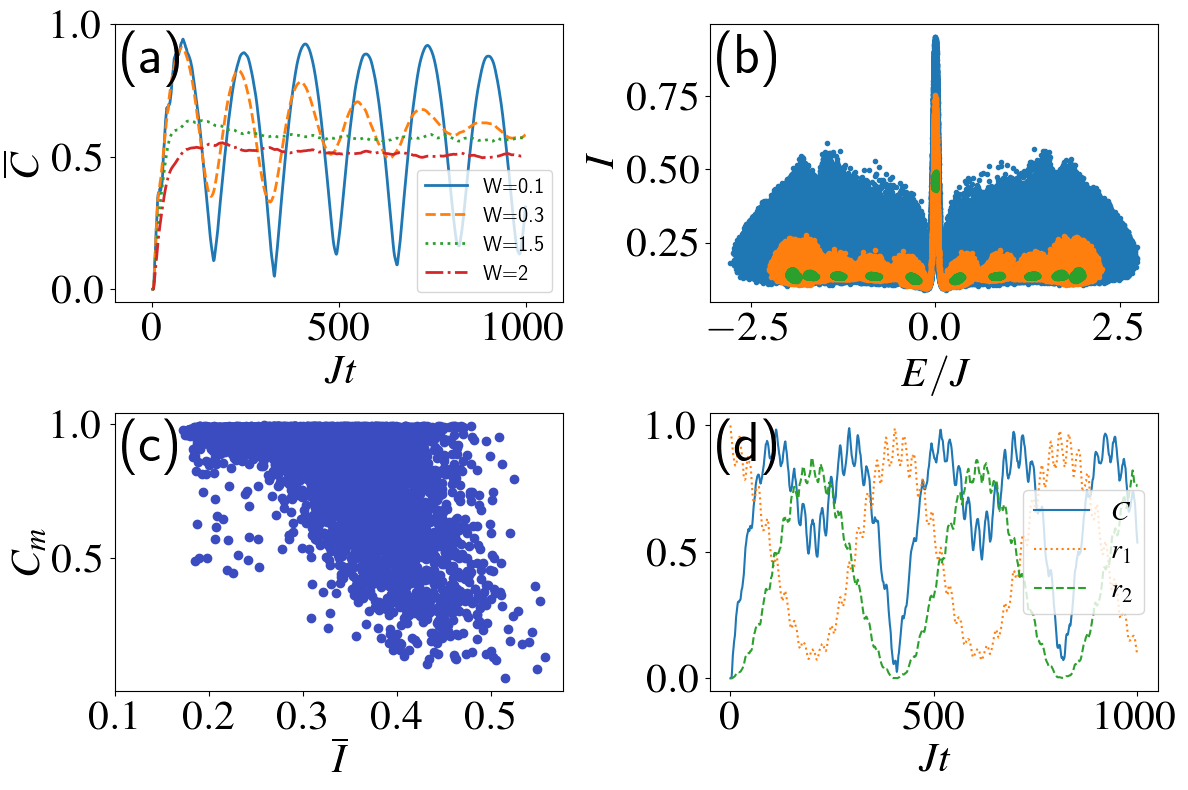}\\
\caption{Effect of disorder on concurrence. (a) Average concurrence over 1000 random realizations for each value of $W$: $W=0.1, 0.3, 1.5, 2$. (b) IPR of eigenstates within the single excitation Hamiltonian versus energy for $W=$ 0.1 (green), 0.4 (orange), 1 (blue). Here, $10^5$ realizations are performed for each $W$. (c) Maximum concurrence versus average IPR for all realizations in panel (a). (d) Concurrence dynamics and return probabilities for atoms $n_{1}$ ($r_1$) and $n_{2}$ ($r_2$) for a particular realization with $W=1$. On-site energies are $\Delta_{c_{i}}/J=$ [$-$0.71, 0.86, 0.19, 0.42, 0.50, $-$0.24, 0.19, $-$0.45, 0.18, $-$1.00] and $\Delta_{n_{1}}=\Delta_{n_{2}}=0$.}
\label{disorder}
\end{figure}

\textit{Effect of disorder.---} We now investigate the effect of disorder in the spin cavity on entanglement generation. In previous calculations, the two atoms were resonant with the on-site energies of the cavity spins, $\Delta_{n_1}=\Delta_{n_2}=\Delta_{c_i}=0$. Now, we assume the on-site energies of the cavity spins are randomly distributed in the interval $\Delta_{c_{i}}/J\in[-W,W]$. Atom $n_{1}$ is initially prepared in state $|1\rangle$, while atom $n_{2}$ and all cavity spins are in the ground state $|0\rangle$, as in the first part of the manuscript. The cavity length is $L=10$. Details of the single-excitation Hamiltonian are in Section I of the SM \cite{sm}. Fig. \ref{disorder}(a) shows the evolution of average concurrence $\overline{C}$ for different disorder strengths $W$. As expected, the average concurrence $\overline{C}$ gradually decreases as $W$ increases.

When studying localization phenomena \cite{PhysRev.109.1492}, it is standard to analyze the inverse participation ratio (IPR), defined as $I_{j}=\sum_{i=1}^{N_{T}}|\beta^{j}_{i}|^4$, $\beta^{j}_{i}$ being the $i$-th component of the $j$-th eigenstate. The IPR measures the spatial spreading of the wavefunction, with higher values indicating more localization. In Fig. \ref{disorder}(b), we plot the IPR of the eigenstate of the single-excitation Hamiltonian versus energy for three disorder strengths. For weak disorder ($W=0.1$, green), most wavefunctions are delocalized except those near the atoms' energy ($\Delta_{n_1}=\Delta_{n_2}=0$), which are more localized to mediate entanglement. As disorder increases, the energy gap closes (green to orange), and states become more localized. However, even at strong disorder ($W=1$), some states remain delocalized (low IPR), which is good for entangling two atoms. In Fig. \ref{disorder}(c), we render maximum concurrence $C_{m}$ versus average IPR ($\overline{I}=\sum{I_j}/N_{T}$), showing a decreasing trend for increasing IPR. Higher concurrence ($C_{m}>0.5$) tends to occur when $\overline{I}<0.35$. Another criterion for generating high entanglement in disordered systems is the effective exchange of return probabilities for atom $n_{i}$, given by $r_{i}=|\langle{00\cdot\cdot\cdot{1_{n_{i}}}\cdot\cdot\cdot00}|\psi(t)\rangle|^2$, where $|\psi(t)\rangle$ is the wavefunction of the system at time $t$. In Fig. \ref{disorder}(d), we show the time evolution of concurrence (blue line) for a disorder realization ($W=1$, $\overline{I}=0.23$), with $r_1$ (orange dotted line) and $r_2$ (green dotted line). Even for strong disorder, maximum entanglement ($\sim 0.99$) occurs when $r_1 \approx r_2$, and minimum when $r_1$ and $r_2$ differ by a large value.

\begin{figure}[htp]
\includegraphics[width=8.5cm]{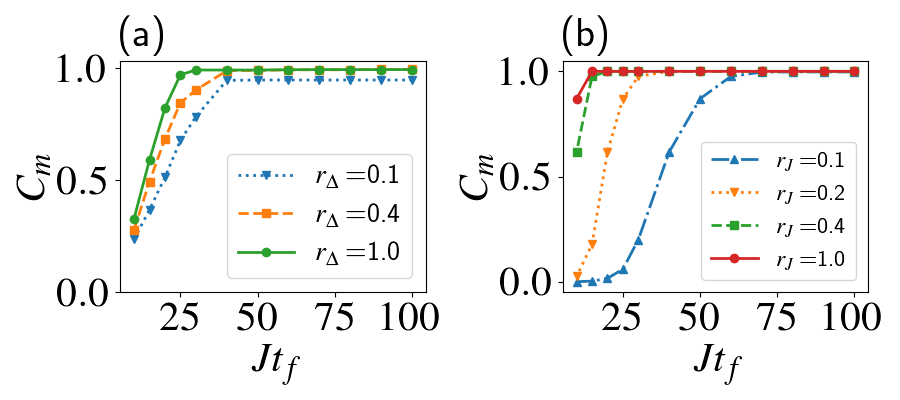}\\
\caption{Fast entanglement generation by engineering on-site energies and hoppings. The cavity length is $L=10$, with atoms located at $\mathcal{L}[n_{1}]=2$ and $\mathcal{L}[n_{2}]=8$. Initially, atom $n_1$ is excited to state $|1\rangle$, while atom $n_2$ is in the ground state $|0\rangle$. All all cavity spins remain in the ground state. Maximum concurrence versus stopping time, $Jt_{f}$, for different restrictions on engineering on-site energies, $r_{\Delta}$, in panel (a), and hoppings, $r_{J}$, in panel (b).}
\label{optimize}
\end{figure}

\textit{Fast entanglement generation.---} In the previous section, we found that disorder affects entanglement generation, and the average IPR can estimate the expected degree of entanglement. Here, we propose fast, high-concurrence entanglement generation by engineering either the on-site energies or hoppings in the cavity. Our goal is to maximize concurrence within a time interval $[0, Jt_f]$, where $Jt_{f}$ is the stopping time in time evolution. When engineering on-site energies, we restrict their values to $[-r_{\Delta},r_{\Delta}]$ by setting $\Delta_{c_{i}}=r_{\Delta}\cos(\theta_{i})$, keeping the hoppings the same as those in Fig. \ref{model}. For tuning hoppings, we restrict them to $[0,r_{J}]$ by setting $J_{c_{i}}(g_{n_{i}})=r_{J}\cos^{2}(\theta_{c_{i}}(\theta_{n_{i}})/2)$, while on-site energies are the same as those in Fig. \ref{model}. Thus, maximum concurrence in $[0,Jt_{f}]$ becomes a function of angles, $C_{m}(\{\theta_{i}\})$ or $C_{m}(\{\theta_{c_i},\theta_{n_i}\})$. In Fig. \ref{optimize}, we show optimization results for both scenarios using the \texttt{Powell} method in \texttt{SciPy}. Compared to the ordered case (Fig. \ref{model}), engineering on-site energies (panel a) or hoppings (panel b) significantly accelerates high-concurrence generation. As $r_{\Delta(J)}$ increases, the two atoms reach high concurrence ($C_{m}>0.95$) within a shorter time $Jt_{f}$ than in the ordered system. A detailed description of the parameters for achieving this improved performance, the parameter tolerance and dissipation effects are provided in Section V of the SM \cite{sm}.

It is important to highlight that compared to the time-dependent optimal control schemes previously proposed \cite{PhysRevLett.103.110501,PhysRevApplied.18.044042,PhysRevA.82.040305}, our platform achieves a very high degree of entanglement without relying on any complex control of the fields and, therefore, is easier to implement experimentally by either using Rydberg atoms as an analog quantum simulator or utilizing the IBM quantum device, IBM QX20 Tokyo \cite{Gheorghiu2020ReducingTC,e25030465,10.1007/978-3-030-52482-1_11}. Both implementation schemes are detailed in Section VI of the SM \cite{sm}.

\textit{Conclusions.---} We have demonstrated that parity and chirality effects play a critical role in entanglement generation between atoms chirally coupled to a spin cavity.
Using perturbation theory and the variational MPS algorithm, we find that small even-sized cavities, which avoid resonant magnonic modes, show higher concurrence and more efficient entanglement generation than odd-sized cavities.
Additionally, while classical driving fields generally reduce concurrence by pushing the atom subsystem out of the single-excitation space where the Bell state resides, their influence is non-monotonic, with concurrence dips corresponding to resonances between the driven atoms and the spin cavity.
Furthermore, when dealing with energetically-disordered spin cavities, we find that the inverse participation ratio of the resulting eigenstates, rather than disorder strength, determines entanglement generation between the two atoms.
Consequently, high concurrence can be observed even in the presence of strong disorder.
The introduction of disorder can significantly expedite entanglement generation (3 to 4 times faster), with high concurrence ($>0.999$) achievable even under strong disorder by fine-tuning the cavity parameters.
Our study showcases the potential of spin cavities as a powerful platform for entanglement generation in quantum systems.

\begin{acknowledgments}
The IHPC A*STAR Team acknowledges support from A*STAR Career Development Award (C210112010), and A*STAR (C230917003, C230917007). The MPS codes that support the findings of this study are available under a reasonable request from the corresponding authors.
\end{acknowledgments}

\clearpage

\onecolumngrid

\section*{Supplemental material: Generation and optimization of entanglement \\ between atoms chirally coupled to spin cavities}

\renewcommand{\thefigure}{S\arabic{figure}}
\renewcommand{\thetable}{S\arabic{table}}

\section{Model Hamiltonian}
\label{model}

\begin{figure}[htbp]
\includegraphics[width=9cm]{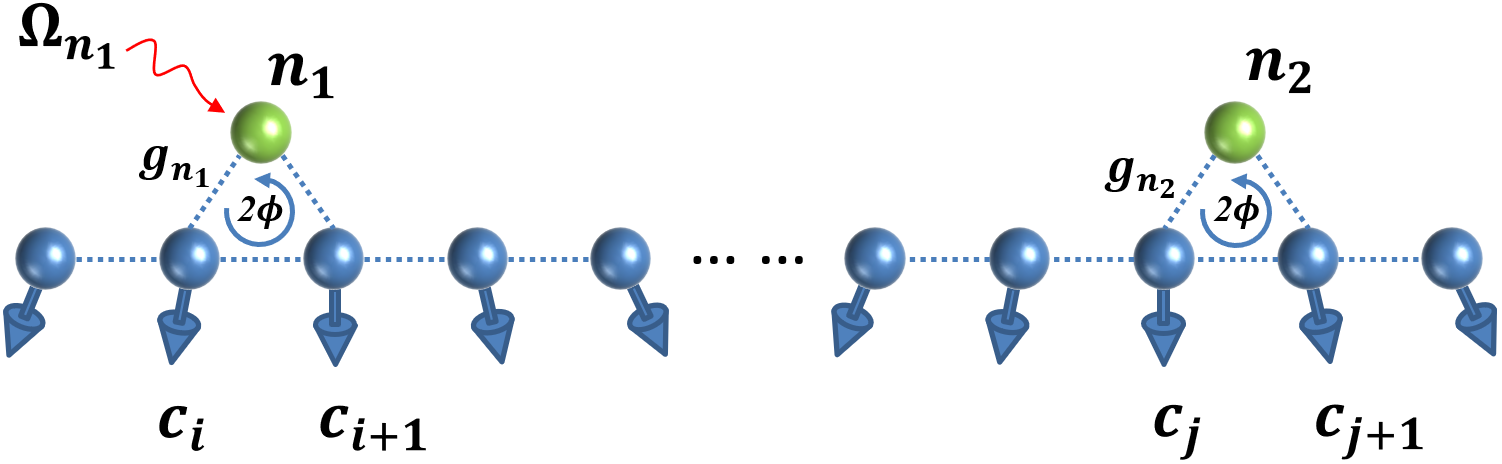}\\
\caption{Schematic of driven atoms coupled to a spin cavity.}\label{model}
\end{figure}

We consider the finite spin chain with open boundary condition as a spin cavity,  where the spin wave is the excitation of the cavity. Two atoms are coupled laterally to the spin cavity.
The Hamiltonian of the spin cavity is given by
\begin{equation}
\label{cavity}
\begin{split}
H_{c}=\sum_{i=1}^{L}\Delta_{c_{i}}\sigma_{c_{i}}^{+}\sigma_{c_{i}}^{-}+\sum_{i=1}^{L-1}J_{c_{i}}(\sigma_{c_{i+1}}^{+}\sigma_{c_{i}}^{-}+\sigma_{c_{i}}^{+}\sigma_{c_{i+1}}^{-}),\\
\end{split}
\end{equation}
where $\sigma_{c_{i}}^{\pm}$ are the rising and lowering operators of the cavity spins, $\Delta_{c_{i}}$ and $J_{c_{i}}$ are the on-site energies and the hopping strengths. The Hamiltonian of the two atoms is given by
\begin{equation}
\label{atoms}
\begin{split}
H_{n}=\sum_{i=1}^{2}\Delta_{n_{i}}\sigma_{n_{i}}^{+}\sigma_{n_{i}}^{-}+\Omega_{n_{i}}(\sigma_{n_{i}}^{+}+\sigma_{n_{i}}^{-}),\\
\end{split}
\end{equation}
where $\sigma_{n_i}^{\pm}$ are the rising and lowering operators of the atoms, $\Delta_{n_i}$ and $\Omega_{n_i}$ are the on-site energies and the driving strengths.
The coupling between atoms and spin cavity is given by
\begin{equation}
\label{couplings}
\begin{split}
H_{g}&=\sum_{i=1}^{2}g_{n_{i}}e^{-i\phi_{n_{i}}}(\sigma_{n_{i}}^{+}\sigma_{c_{\mathcal{L}[n_{i}]}}^{-}+\sigma_{c_{\mathcal{R}[n_{i}]}}^{+}\sigma_{n_{i}}^{-})+\text{h.c.},\\
\end{split}
\end{equation}
where $g_{n_{i}}$ and $\phi_{n_{i}}$ are the coupling strength and hopping phase, and $\mathcal{L}[n_{i}]$ and $\mathcal{R}[n_{i}]$ represent the cavity spins to the left and right of atom $n_{i}$.
These can be used to label the positions of the atoms along the cavity.

In the following discussion, we study the case where the atomic system is resonant with the spin cavity, $\Delta_{c_i}=\Delta_{n_i}=\Delta=0$ and the hopping strength of the spin cavity is homogeneous, $J_{c_i}=J$.

\subsection{Dispersion of spin cavity}

We can diagonalize the spin cavity in Eq. (\ref{cavity}) to obtain the dispersion of the spin cavity. It is found that the Hamiltonian can be diagonalized by the unitary transformation,
\begin{equation}
\label{unitarytransform}
\begin{split}
\sigma_{c_i}^{-}=\sqrt{\frac{2}{L+1}}\sum_{k=1}^{L}\sin{\frac{\pi k i}{L+1}}\eta_{c,k}.
\end{split}
\end{equation}
After the transformation, the Hamiltonian can be recast into
\begin{equation}
\begin{split}
H_{c}=\sum_{k=1}^{L}\epsilon_{c,k}\eta_{c,k}^{\dag}\eta_{c,k},
\end{split}
\end{equation}
where the dispersion of spin cavity is
\begin{equation}
\label{dispersion}
\begin{split}
\epsilon_{c,k}=\Delta+2J\cos\frac{\pi k}{L+1}.
\end{split}
\end{equation}
Note that by the orthogonality relation,
\begin{equation}
\begin{split}
\sum_{j=1}^{L}\sin{\frac{\pi kj}{L+1}}\sin{\frac{\pi lj}{L+1}}=\frac{L+1}{2}\delta_{kl},
\end{split}
\end{equation}
the eigenmode of the spin cavity can be written as
\begin{equation}
\begin{split}
\eta_{c,k}=\sqrt{\frac{2}{L+1}}\sum_{j=1}^{L}\sin{\frac{\pi k j}{L+1}}\sigma_{c_j}^{-}.
\end{split}
\end{equation}

From the dispersion relation of the spin cavity in Eq. (\ref{dispersion}), we observe that for an odd-sized cavity, the eigenmode $\eta_{k=\frac{N+1}{2}}$ has an energy of $\Delta$, which is resonant with the atomic system. In contrast, for an even-sized cavity, there is no cavity mode resonant with the atomic system. The leakage of excitation from the atomic system to the resonant eigenmode of the spin cavity results in low concurrence in the odd-sized cavity.

Finally, we can employ second-order perturbation theory (as described in Section III of this Supplemental Material) to derive an effective Hamiltonian for the atomic subsystem in both even- and odd-sized cavities. We begin by recasting the coupling term into the space spanned by the eigenmodes of the spin cavity and the atoms.
\begin{equation}
\begin{split}
H_{g}&=\sqrt{\frac{2}{L+1}}\sum_{k=1}^{L}\sin{\frac{\pi k i}{L+1}}(g_{n_{1}}e^{-i\phi_{n_{1}}}\sigma_{n_{1}}^{+}\eta_{c,k}+g_{n_{1}}e^{i\phi_{n_{1}}}\eta_{c,k}^{\dag}\sigma_{n_{1}}^{-})\\
&+\sqrt{\frac{2}{L+1}}\sum_{k=1}^{L}\sin{\frac{\pi k (i+1)}{L+1}}(g_{n_{1}}e^{i\phi_{n_{1}}}\sigma_{n_{1}}^{+}\eta_{c,k}+g_{n_{1}}e^{-i\phi_{n_{1}}}\eta_{c,k}^{\dag}\sigma_{n_{1}}^{-})\\
&+\sqrt{\frac{2}{L+1}}\sum_{k=1}^{L}\sin{\frac{\pi k j}{L+1}}(g_{n_{2}}e^{-i\phi_{n_{2}}}\sigma_{n_{2}}^{+}\eta_{c,k}+g_{n_{2}}e^{i\phi_{n_{2}}}\eta_{c,k}^{\dag}\sigma_{n_{2}}^{-})\\
&+\sqrt{\frac{2}{L+1}}\sum_{k=1}^{L}\sin{\frac{\pi k (j+1)}{L+1}}(g_{n_{2}}e^{i\phi_{n_{2}}}\sigma_{n_{2}}^{+}\eta_{c,k}+g_{n_{2}}e^{-i\phi_{n_{2}}}\eta_{c,k}^{\dag}\sigma_{n_{2}}^{-}).\\
\end{split}
\end{equation}

\subsection{Relation between chirality and hopping phase}

We can impose periodic boundary conditions on the cavity Hamiltonian Eq. (\ref{cavity}) to study the propagation direction of excitations with respect to the hopping phase. By the Fourier transform, $\sigma_{c_i}^{+}=\frac{1}{\sqrt{L}}\sum_{K}e^{-iKR_{i}}\sigma_{c,K}^{+}$, The cavity Hamiltonian in the momentum space is given by $H_{c}=\sum_{K}\varepsilon_{K}\sigma_{c,K}^{+}\sigma_{c,K}^{-}$, where the dispersion relation is $\varepsilon_{K}=\Delta+2J\cos{Ka}$, $K=\frac{2\pi j}{La}\in[-\pi/a,\pi/a)$ and $R_{i}=ia$. Let's consider an atom couples to the spin cavity. In momentum space, the coupling between the atom and the spin cavity can be recast into
\begin{equation}
\begin{split}
H_{g}&=\frac{2g_{n_1}}{\sqrt{L}}\sum_{K}\cos{(Ka/2+\phi_{n_1})}[e^{-iK(R_{i}+a/2)}\sigma_{n_1}^{-}\sigma_{c,K}^{+}+e^{iK(R_{i}+a/2)}\sigma_{n_1}^{+}\sigma_{c,K}^{-}].\\
\end{split}
\end{equation}

\begin{figure}[htp]
\centering
\includegraphics[width=16cm]{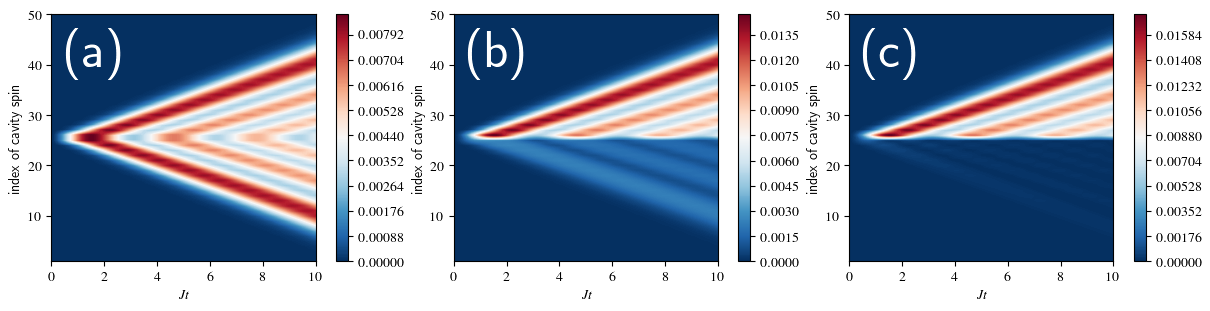}\\
\caption{Excitation versus time for different hopping phases, (a) $\phi_{n_1}=0$, (b) $\phi_{n_1}=\pi/8$ and (c) $\phi_{n_1}=\pi/4$. The atom is positioned in the middle of the spin cavity with length $L=50$ and is initially excited in the state $|1\rangle$.}
\label{propogation}
\end{figure}

According to the arguments in Ref. \cite{PhysRevA.93.062104}, magnons with $K>0$ ($K<0$) have negative (positive) group velocity $v_{K}=\partial{\varepsilon_{K}}/\partial{K}$, and thus propagate to the left (right) along the spin cavity. In the momentum space, the coupling becomes $K$-dependent, $g_{K}=\frac{2g_{n_1}}{\sqrt{L}}\cos{(Ka/2+\phi_{n_1})}$. Importantly, the hopping phase $\phi_{n_1}$ renders this coupling asymmetric in $K$ and thus makes it chiral.
For $\phi_{n_1}=\pi/4$, we find that all magnonic modes moving to the right couple stronger than the ones to the left.
Additionally, we can illustrate this relation by numerically showing the propagation of excitation as a function of the hopping phase, specifically for $\phi_{n_1}=0$, $\phi_{n_1}=\pi/8$ and $\phi_{n_1}=\pi/4$ as shown in Fig. (\ref{propogation}). We can observe a gradual change in the propagation direction as the hopping phase increases from $0$ to $\pi/4$.

\subsection{Maximal concurrence at the first occurrence and corresponding optimal time for different system size}

\begin{figure}[htp]
\includegraphics[width=16cm]{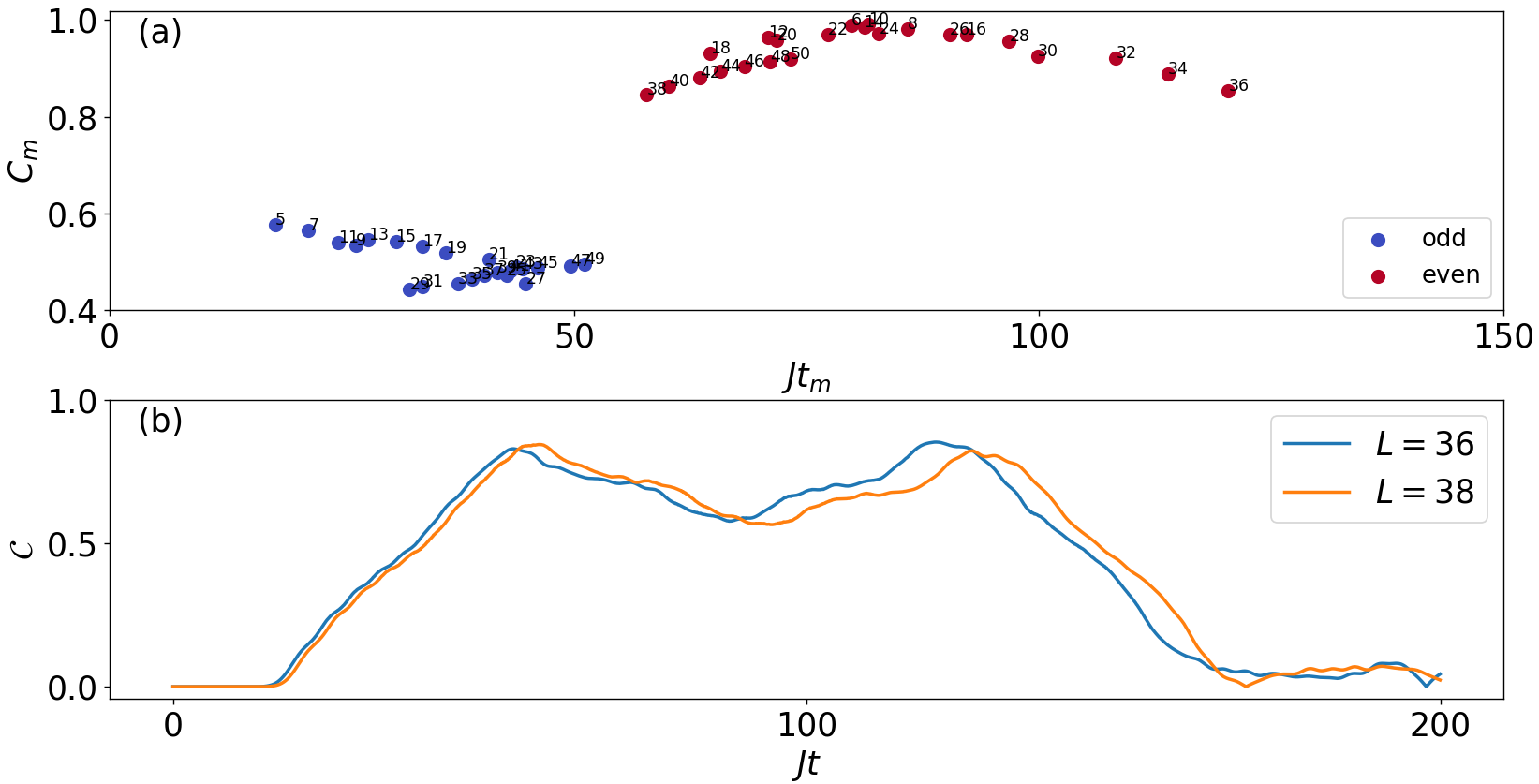}\\
\caption{(a) Maximum concurrence $C_{m}$ at the first occurrence and the corresponding optimal time $Jt_{m}$ for both odd- and even-sized cavities. The data points labeled with integers represent the cavity length $L$. (b) Time evolution of concurrence $\mathcal{C}$ for cavity lengths $L = 36$ (blue line) and $L = 38$ (orange line).}
\label{figs_label}
\end{figure}

In Fig. \ref{figs_label}(a), we show the maximum concurrence $C_{m}$ at the first occurrence and the corresponding optimal time, $Jt_{m}$, for both odd- and even-sized cavities. The data points labeled with integers represent the cavity length $L$.
It is observed that the even-sized cavities (red points) consistently reach higher concurrence values ($C_{m}>0.8$) over shorter time scales compared to the odd-sized cavities (blue points), where \(C_{m}\) remains below $0.6$.
This discrepancy underscores the influence of cavity size parity on the entanglement dynamics, with even-sized cavities exhibiting more favorable conditions for rapid and strong entanglement generation.
The maximum $C_{m}$ of approximately $0.99$ occurs for $L=10$.
Furthermore, for cavities with the same parity, $C_{m}$ tends to decrease as the cavity length increases.
This occurs because the increasing number of cavity modes that become near-resonant with the atomic modes at energy $\Delta$ leads to greater leakage into the spin cavity.

In Fig. \ref{figs_label}(a), we observe that very similar systems can exhibit markedly different behaviors in terms of concurrence, such as those with cavity lengths $L=36$ and $L=38$.
To investigate the underlying reasons, we plot the time evolution of concurrence in Fig. \ref{figs_label}(b). The analysis reveals that this difference arises from the imbalance between the two main concurrence peaks and the transition from dominance of one peak to the other as the cavity length increases from 36 to 38.

\subsection{Single-excitation Hamiltonian for the spin cavity with length $L=10$}

In this section, we provide the explicit matrix form of the single-excitation Hamiltonian for the spin cavity of length $L=10$, as discussed in the main text. This Hamiltonian governs the dynamics of the system when restricted to the subspace with a single excitation, either in the cavity spins or in one of the atoms.

\begin{equation}
\begin{split}
H_{\text{1-exc}}=&\left[\begin{array}{*{20}cccccccccccc}
{\Delta_{c_{1}}} & {J_{c_{1}}} & {0} & {0} & {0} & {0} & {0} & {0} & {0} & {0} & {0} & {0}\\
{J_{c_{1}}} & {\Delta_{c_{2}}} & {g_{n_{1}}e^{i\phi_{n_{1}}}} & {J_{c_{2}}} & {0} & {0} & {0} & {0} & {0} & {0} & {0} & {0}\\
{0} & {g_{n_{1}}e^{-i\phi_{n_{1}}}} & {\Delta_{n_{1}}} & {g_{n_{1}}e^{i\phi_{n_{1}}}} & {0} & {0} & {0} & {0} & {0} & {0} & {0} & {0}\\
{0} & {J_{c_{2}}} & {g_{n_{1}}e^{-i\phi_{n_{1}}}} & {\Delta_{c_{3}}} & {J_{c_{3}}} & {0} & {0} & {0} & {0} & {0} & {0} & {0}\\
{0} & {0} & {0} & {J_{c_{3}}} & {\Delta_{c_{4}}} & {J_{c_{4}}} & {0} & {0} & {0} & {0} & {0} & {0}\\
{0} & {0} & {0} & {0} & {J_{c_{4}}} & {\Delta_{c_{5}}} & {J_{c_{5}}} & {0} & {0} & {0} & {0} & {0}\\
{0} & {0} & {0} & {0} & {0} & {J_{c_{5}}} & {\Delta_{c_{6}}} & {J_{c_{6}}} & {0} & {0} & {0} & {0}\\
{0} & {0} & {0} & {0} & {0} & {0} & {J_{c_{6}}} & {\Delta_{c_{7}}} & {J_{c_{7}}} & {0} & {0} & {0}\\
{0} & {0} & {0} & {0} & {0} & {0} & {0} & {J_{c_{7}}} & {\Delta_{c_{8}}} & {g_{n_{2}}e^{i\phi_{n_{2}}}} & {J_{c_{8}}} & {0}\\
{0} & {0} & {0} & {0} & {0} & {0} & {0} & {0} & {g_{n_{2}}e^{-i\phi_{n_{2}}}} & {\Delta_{n_{2}}} & {g_{n_{2}}e^{i\phi_{n_{2}}}} & {0}\\
{0} & {0} & {0} & {0} & {0} & {0} & {0} & {0} & {J_{c_{8}}} & {g_{n_{2}}e^{-i\phi_{n_{2}}}} & {\Delta_{c_{9}}} & {J_{c_{9}}}\\
{0} & {0} & {0} & {0} & {0} & {0} & {0} & {0} & {0} & {0} & {J_{c_{9}}} & {\Delta_{c_{10}}}\\
\end{array}\right]\\
\end{split}
\end{equation}

\section{Effects of dissipation on the concurrence}

\begin{figure}[htp]
\includegraphics[width=12cm]{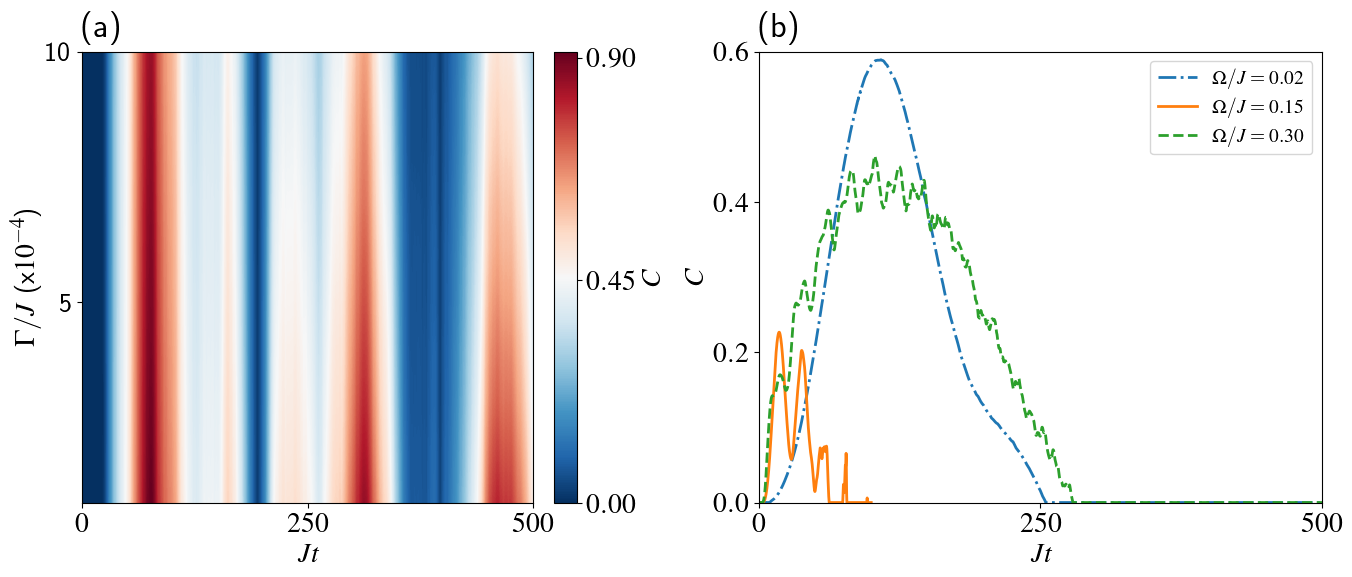}\\
\caption{Effects of dissipation on the concurrence of two atoms within a spin cavity. (a) The cavity length is denoted by $L=50$, with two atoms $N=2$ in the cavity. The on-site energy $\Delta$ and driving strength $\Omega_{n_{i}}$ are set to zero. Initially, atom $n_{1}$ is prepared in the excited state $|1\rangle$, and atom $n_{2}$ is in the ground state $|0\rangle$. The hopping amplitude in the spin cavity is $J_{c_{i}}=J=1$, and the coupling between the atoms and the cavity is $g_{n_{i}}e^{-i\phi_{n_{i}}}{\equiv}ge^{-i\phi}$, where $g/J=0.1$. The dissipation $\Gamma/J$ is from $10^{-4}$ to $10^{-3}$. The bond dimension $D=10$ is used for the MPS simulation. (b) The dissipation rate of $\Gamma_{c_{i}}/J = \Gamma_{n_{i}}/J = 0.005$, with driving strengths $\Omega_{n_{1}}/J = 0.02, 0.15, 0.3$. Initially, the atoms and the cavity spins are in the ground state $|0\rangle$. The bond dimension is $D=30$.}
\label{dissipation}
\end{figure}

The dissipation from the system to the environment can be described by the Lindblad master equation (with $\hbar = 1$),
\begin{equation}
\label{Lindblad}
\begin{split}
\frac{d\rho}{dt}=i[\rho,H]+\sum_{i=1}^{L}\frac{\Gamma_{c_{i}}}{2}\mathcal{D}[\sigma_{c_{i}}^{-}]\rho+\sum_{i=1}^{2}\frac{\Gamma_{n_{i}}}{2}\mathcal{D}[\sigma_{n_{i}}^{-}]\rho,\\
\end{split}
\end{equation}
where $\rho$ is the density matrix of the system, and $\mathcal{D}[\sigma_{c_{i}(n_{i})}^{-}]\rho=2\sigma_{c_{i}(n_{i})}^{-}\rho\sigma_{c_{i}(n_{i})}^{+}-\sigma_{c_{i}(n_{i})}^{+}\sigma_{c_{i}(n_{i})}^{-}\rho-\rho\sigma_{c_{i}(n_{i})}^{+}\sigma_{c_{i}(n_{i})}^{-}$ represents the Lindblad dissipator with decay rates $\Gamma_{c_{i}(n_{i})}$ for the spin cavity and atoms, respectively.

In Fig. \ref{dissipation}(a), we study the effect of dissipation on concurrence and observe, as expected, that concurrence decreases with increasing dissipation strength. Under dissipation, the concurrence peaks gradually diminish over time. In Fig. \ref{dissipation}(b), we explore the influence of dissipation in the case of a driven atom. We find non-monotonic behavior for the maximal concurrence at a dissipation rate of $\Gamma_{c_{i}}/J = \Gamma_{n_{i}}/J = 0.005$, with driving strengths $\Omega_{n_{1}}/J = 0.02, 0.15, 0.3$.

\section{Second-order perturbation theory}
\label{2ndperturb}

In this section, we will use second order perturbation theory to study the parity effect related to cavity length and atom distance. Generally, in an odd-sized cavity, there is an eigenmode that resonates with the atoms, resulting in low entanglement between the two atoms. Conversely, in an even-sized cavity, no cavity mode resonates with the atomic system, leading to high entanglement.

\subsection{Single-excitation subspace}

We first consider the scenario of two atoms without external driving, $\Omega_{n_i}=0$. The initial state is prepared with atom $n_{1}$ in an excited state and all others in the ground state, $\psi(0)=|1_{n_1}0_{n_2}\rangle\otimes|0_{c_1}\cdots0_{c_L}\rangle$, which is equivalent to $\psi(0)=|1_{n_1}0_{n_2}\rangle\otimes|0_{\eta_{c,1}}\cdots0_{\eta_{c,L}}\rangle$. In this case, the relevant dynamics can be restricted to the single-excitation subspace. Assuming the coupling strength between the spin cavity and the atoms is smaller than the energy scale of the cavity, $g_{n_1}, g_{n_2} \leq J$, we can use perturbation theory to derive the effective Hamiltonian of the atomic subsystem for both even- and odd-sized cavities.

\subsubsection{Even size}

For a cavity of even size, based on the dispersion in Eq. (\ref{dispersion}), there is no eigenmode in the cavity that is resonant with the atomic system. Therefore, we project the Hamiltonian into the subspace spanned by the eigenmodes of the two-atom subsystem. The projectors for the degenerate subspace (atomic system) and the rest of the single-excitation space are given by
\begin{equation}
\begin{split}
P&=\sigma_{n_1}^{+}|{0}\rangle\langle{0}|\sigma_{n_1}^{-}+\sigma_{n_2}^{+}|{0}\rangle\langle{0}|\sigma_{n_2}^{-},\\
Q&=\sum_{k=1}^{L}\eta_{c,k}^{\dag}|{0}\rangle\langle{0}|\eta_{c,k}.\\
\end{split}
\end{equation}
The unperturbed Hamiltonian $H_{0}=H_{n}+H_{c}$ can be divided by
\begin{equation}
\begin{split}
H_{P}^{(0)}&=\sum_{i=1}^{2}\Delta\sigma_{n_i}^{+}\sigma_{n_i}^{-},\\
H_{Q}^{(0)}&=\sum_{k=1}^{L}\epsilon_{c,k}\eta_{c,k}^{\dag}\eta_{c,k}.\\
\end{split}
\end{equation}
Since the perturbation Hamiltonian $H_{g}$ (atom-cavity coupling) is off-diagonal, the first order correction $H_{P}^{(1)}=PH_{g}P$ is zero. The second order correction is
\begin{equation}
\begin{split}
H_{P}^{(2)}=-PH_{g}Q\frac{1}{H_{0}-E_{P}^{(0)}}QH_{g}P,\\
\end{split}
\end{equation}
where $E_{P}^{(0)}=\Delta$ is the degenerate energy in $P$ subspace.
Define the eigenmodes in $P$ subspace as $|n_{1}\rangle=\sigma_{n_1}^{+}|0\rangle$ and $|n_{2}\rangle=\sigma_{n_2}^{+}|0\rangle$, the matrix element for $H_{P}^{(2)}$ is
\begin{equation}
\begin{split}
\langle{n_{\alpha}}|H_{P}^{(2)}|{n_{\beta}}\rangle=-\sum_{k=1}^{L}\frac{\langle{0}|\sigma_{n_{\alpha}}^{-}|H_{g}|\eta_{c,k}^{\dag}|{0}\rangle\langle{0}|\eta_{c,k}|H_{g}|\sigma_{n_{\beta}}^{+}|{0}\rangle}{\epsilon_{c,k}-\Delta},\\
\end{split}
\end{equation}
where $\alpha,\beta=1,2$. It is straightforward to obtain that
\begin{equation}
\begin{split}
&\langle{0}|\sigma_{n_{\alpha}}^{-}|H_{g}|\eta_{c,k}^{\dag}|{0}\rangle=\\
&\sqrt{\frac{2}{L+1}}\left[g_{n_{1}}(e^{-i\phi_{n_{1}}}\sin{\frac{\pi k i}{L+1}}+e^{i\phi_{n_{1}}}\sin{\frac{\pi k (i+1)}{L+1}})\delta_{1,\alpha}+g_{n_{2}}(e^{-i\phi_{n_{2}}}\sin{\frac{\pi k j}{L+1}}+e^{i\phi_{n_{2}}}\sin{\frac{\pi k (j+1)}{L+1}})\delta_{2,\alpha}\right].\\
\end{split}
\end{equation}

Therefore, the matrix elements for $H_{P}^{(2)}$ are
\begin{equation}
\begin{split}
\langle{n_{1}}|H_{P}^{(2)}|{n_{1}}\rangle&=-\frac{g_{n_{1}}^2}{J(L+1)}\sum_{k=1}^{L}\frac{(e^{-i\phi_{n_{1}}}\sin{\frac{\pi k i}{L+1}}+e^{i\phi_{n_{1}}}\sin{\frac{\pi k (i+1)}{L+1}})(e^{i\phi_{n_{1}}}\sin{\frac{\pi k i}{L+1}}+e^{-i\phi_{n_{1}}}\sin{\frac{\pi k (i+1)}{L+1}})}{\cos\frac{\pi k}{L+1}},\\
&=-\frac{g_{n_{1}}^{2}}{J}(1-(-1)^{i})\cos{2\phi_{n_{1}}},\\
\langle{n_{2}}|H_{P}^{(2)}|{n_{2}}\rangle&=-\frac{g_{n_{2}}^2}{J(L+1)}\sum_{k=1}^{L}\frac{(e^{-i\phi_{n_{2}}}\sin{\frac{\pi k j}{L+1}}+e^{i\phi_{n_{2}}}\sin{\frac{\pi k (j+1)}{L+1}})(e^{i\phi_{n_{2}}}\sin{\frac{\pi k j}{L+1}}+e^{-i\phi_{n_{2}}}\sin{\frac{\pi k (j+1)}{L+1}})}{\cos\frac{\pi k}{L+1}},\\
&=-\frac{g_{n_{2}}^{2}}{J}(1-(-1)^{j})\cos{2\phi_{n_{2}}},\\
\langle{n_{1}}|H_{P}^{(2)}|{n_{2}}\rangle&=-\frac{g_{n_{1}}g_{n_{2}}}{J(L+1)}\sum_{k=1}^{L}\frac{(e^{-i\phi_{n_{1}}}\sin{\frac{\pi k i}{L+1}}+e^{i\phi_{n_{1}}}\sin{\frac{\pi k (i+1)}{L+1}})(e^{i\phi_{n_{2}}}\sin{\frac{\pi k j}{L+1}}+e^{-i\phi_{n_{2}}}\sin{\frac{\pi k (j+1)}{L+1}})}{\cos\frac{\pi k}{L+1}},\\
&=\frac{g_{n_{1}}g_{n_{2}}}{J}(\cos{\frac{(i+j)\pi}{2}}+\sin{\frac{(i-j)\pi}{2}})e^{i[(-1)^{i}\phi_{n_{1}}+(-1)^{j}\phi_{n_{2}}]},\\
\langle{n_{2}}|H_{P}^{(2)}|{n_{1}}\rangle&=-\frac{g_{n_{1}}g_{n_{2}}}{J(L+1)}\sum_{k=1}^{L}\frac{(e^{-i\phi_{n_{2}}}\sin{\frac{\pi k j}{L+1}}+e^{i\phi_{n_{2}}}\sin{\frac{\pi k (j+1)}{L+1}})(e^{i\phi_{n_{1}}}\sin{\frac{\pi k i}{L+1}}+e^{-i\phi_{n_{1}}}\sin{\frac{\pi k (i+1)}{L+1}})}{\cos\frac{\pi k}{L+1}},\\
&=\frac{g_{n_{1}}g_{n_{2}}}{J}(\cos{\frac{(i+j)\pi}{2}}+\sin{\frac{(i-j)\pi}{2}})e^{-i[(-1)^{i}\phi_{n_{1}}+(-1)^{j}\phi_{n_{2}}]},\\
\end{split}
\end{equation}
where $i<j$. Therefore, the effective Hamiltonian in $P$ subspace up to the second order is
\begin{equation}
\label{2ndcorr}
\begin{split}
H_{\text{eff}}=&\left[\begin{array}{cc}
{\Delta-\frac{g_{n_{1}}^{2}}{J}(1-(-1)^{i})\cos{2\phi_{n_{1}}}} & {\frac{g_{n_{1}}g_{n_{2}}}{J}(\cos{\frac{(i+j)\pi}{2}}+\sin{\frac{(i-j)\pi}{2}})e^{i[(-1)^{i}\phi_{n_{1}}+(-1)^{j}\phi_{n_{2}}]}}\\
{\frac{g_{n_{1}}g_{n_{2}}}{J}(\cos{\frac{(i+j)\pi}{2}}+\sin{\frac{(i-j)\pi}{2}})e^{-i[(-1)^{i}\phi_{n_{1}}+(-1)^{j}\phi_{n_{2}}]}} & {\Delta-\frac{g_{n_{2}}^{2}}{J}(1-(-1)^{j})\cos{2\phi_{n_{2}}}}\\
\end{array}\right].
\end{split}
\end{equation}
We find that there is an indirect coupling between the two atoms, which is mediated by the spin cavity. There is also a Lamb shift $-\frac{g_{n_{i}}^{2}}{J}(1-(-1)^{i})\cos{2\phi_{n_{i}}}$ for the atoms induced by the coupling to the spin cavity. The strength of indirect coupling is at the magnitude of $\sim{g_{n_1}g_{n_2}/J}$, which is small for weak coupling strength (for example, $g_{n_1}=g_{n_2}=0.1J$ in the main text).

It is found that there are four possibilities for the positions of two atoms based on the parity. Here we assume $\Delta=0$, $g_{n_1}=g_{n_2}=g$ and $\phi_{n_1}=\phi_{n_2}=\phi$.

\underline{Case 1: $i=2m_{1},j=2m_{2}$:}
The effective Hamiltonian is
\begin{equation}
\begin{split}
H_{\text{eff}}=&\left[\begin{array}{*{20}cc}
{0} & {\frac{g^{2}}{J}e^{i[2\phi-(m_{1}+m_{2})\pi]}}\\
{\frac{g^{2}}{J}e^{-i[2\phi-(m_{1}+m_{2})\pi]}} & {0}\\
\end{array}\right].\\
\end{split}
\end{equation}
For the initial state $\psi(0)=[1,0]^{\text{T}}$, the wavefunction at time $t$ is
\begin{equation}
\begin{split}
\psi(t)=\left[\begin{array}{*{20}c}
{\cos{\frac{g^{2}}{J}t}}\\
{-ie^{-i[2\phi-(m_{1}+m_{2})\pi]}\sin{\frac{g^{2}}{J}t}}\\
\end{array}\right].\\
\end{split}
\end{equation}
Therefore, the concurrence is
\begin{equation}
\label{concurr1}
\begin{split}
C(t)=|\sin{\frac{2g^{2}}{J}t}|.
\end{split}
\end{equation}

\underline{Case 2: $i=2m_{1}+1,j=2m_{2}+1$:}
The effective Hamiltonian is
\begin{equation}
\begin{split}
H_{\text{eff}}=&\left[\begin{array}{*{20}cc}
{-\frac{2g^{2}}{J}\cos{2\phi}} & {\frac{g^{2}}{J}e^{-i[2\phi-(m_{1}+m_{2}+1)\pi]}}\\
{\frac{g^{2}}{J}e^{i[2\phi-(m_{1}+m_{2}+1)\pi]}} & {-\frac{2g^{2}}{J}\cos{2\phi}}\\
\end{array}\right].\\
\end{split}
\end{equation}
For the initial state $\psi(0)=[1,0]^{\text{T}}$, the wavefunction at time $t$ is
\begin{equation}
\begin{split}
\psi(t)=e^{it\frac{2g^{2}}{J}\cos{2\phi}}\left[\begin{array}{*{20}c}
{\cos{\frac{g^{2}}{J}t}}\\
{-ie^{i[2\phi-(m_{1}+m_{2}+1)\pi]}\sin{\frac{g^{2}}{J}t}}\\
\end{array}\right].\\
\end{split}
\end{equation}
Therefore, the concurrence is
\begin{equation}
\begin{split}
C(t)=|\sin{\frac{2g^{2}}{J}t}|.
\end{split}
\end{equation}

\underline{Case 3: $i=2m_{1},j=2m_{2}+1$:}
The effective Hamiltonian is
\begin{equation}
\begin{split}
H_{\text{eff}}=&\left[\begin{array}{*{20}cc}
{0} & {(-1)^{m_{1}-m_{2}+1}\frac{g^{2}}{J}}\\
{(-1)^{m_{1}-m_{2}+1}\frac{g^{2}}{J}} & {-\frac{2g^{2}}{J}\cos{2\phi}}\\
\end{array}\right].\\
\end{split}
\end{equation}
For the initial state $\psi(0)=[1,0]^{\text{T}}$, the wavefunction at time $t$ is
\begin{equation}
\begin{split}
\psi(t)=\left[\begin{array}{*{20}c}
{-i\frac{\cos{2\phi}}{\sqrt{1+\cos^{2}{2\phi}}}\sin{\frac{g^{2}\sqrt{1+\cos^{2}{2\phi}}t}{J}}+\cos{\frac{g^{2}\sqrt{1+\cos^{2}{2\phi}}t}{J}}}\\
{i(-1)^{m_{1}-m_{2}}\frac{1}{\sqrt{1+\cos^{2}{2\phi}}}\sin{\frac{g^{2}\sqrt{1+\cos^{2}{2\phi}}t}{J}}}\\
\end{array}\right].\\
\end{split}
\end{equation}
Therefore, the concurrence is
\begin{equation}
\label{concurr2}
\begin{split}
C(t)=2\sqrt{\left(1-\frac{\sin^{2}{\frac{g^{2}\sqrt{1+\cos^{2}{2\phi}}t}{J}}}{1+\cos^{2}{2\phi}}\right)\frac{\sin^{2}{\frac{g^{2}\sqrt{1+\cos^{2}{2\phi}}t}{J}}}{1+\cos^{2}{2\phi}}}.
\end{split}
\end{equation}

\underline{Case 4: $i=2m_{1}+1,j=2m_{2}$:}
The effective Hamiltonian is
\begin{equation}
\begin{split}
H_{\text{eff}}=&\left[\begin{array}{*{20}cc}
{-\frac{2g^{2}}{J}\cos{2\phi}} & {(-1)^{m_{1}-m_{2}}\frac{g^{2}}{J}}\\
{(-1)^{m_{1}-m_{2}}\frac{g^{2}}{J}} & {0}\\
\end{array}\right].\\
\end{split}
\end{equation}
For the initial state $\psi(0)=[1,0]^{\text{T}}$, the wavefunction at time $t$ is
\begin{equation}
\begin{split}
\psi(t)=\left[\begin{array}{*{20}c}
{i\frac{\cos{2\phi}}{\sqrt{1+\cos^{2}{2\phi}}}\sin{\frac{g^{2}\sqrt{1+\cos^{2}{2\phi}}t}{J}}+\cos{\frac{g^{2}\sqrt{1+\cos^{2}{2\phi}}t}{J}}}\\
{i(-1)^{m_{1}-m_{2}+1}\frac{1}{\sqrt{1+\cos^{2}{2\phi}}}\sin{\frac{g^{2}\sqrt{1+\cos^{2}{2\phi}}t}{J}}}\\
\end{array}\right].\\
\end{split}
\end{equation}
Therefore, the concurrence is
\begin{equation}
\begin{split}
C(t)=2\sqrt{\left(1-\frac{\sin^{2}{\frac{g^{2}\sqrt{1+\cos^{2}{2\phi}}t}{J}}}{1+\cos^{2}{2\phi}}\right)\frac{\sin^{2}{\frac{g^{2}\sqrt{1+\cos^{2}{2\phi}}t}{J}}}{1+\cos^{2}{2\phi}}}.
\end{split}
\end{equation}
It is straightforward to see that for Case 3 and 4, the concurrence $C(t)\le1$ with equality when
\begin{equation}
\begin{split}
\frac{\sin^{2}{\frac{g^{2}\sqrt{1+\cos^{2}{2\phi}}t}{J}}}{1+\cos^{2}{2\phi}}=\frac{1}{2}.
\end{split}
\end{equation}
Therefore, the maximum concurrence happens when
\begin{equation}
\begin{split}
T_{C}=\frac{J}{2g^{2}}\frac{(2k+1)\pi\mp\arccos{(\cos^{2}{2\phi})}}{\sqrt{1+\cos^{2}{2\phi}}}, (k=0,1,2,\cdots).
\end{split}
\end{equation}

\subsubsection{Odd size}

For the odd-sized spin cavity, the eigenmode $\eta_{\Delta}\equiv\eta_{k=\frac{L+1}{2}}$ has an energy of $\Delta$, which is resonant with the atomic system.
Therefore, we project the Hamiltonian into the subspace spanned by the two atomic states and the cavity mode $\eta_{\Delta}$. The projectors for the degenerate subspace and the rest of the single-excitation subspace are
\begin{equation}
\begin{split}
P&=\sigma_{n_1}^{+}|{0}\rangle\langle{0}|\sigma_{n_1}^{-}+\sigma_{n_2}^{+}|{0}\rangle\langle{0}|\sigma_{n_2}^{-}+\eta_{\Delta}^{\dag}|{0}\rangle\langle{0}|\eta_{\Delta},\\
Q&=\sum_{k=1,\ne\frac{L+1}{2}}^{L}\eta_{c,k}^{\dag}|{0}\rangle\langle{0}|\eta_{c,k}.\\
\end{split}
\end{equation}
The unperturbed Hamiltonian $H_{0}=H_{n}+H_{c}$ can be divided by
\begin{equation}
\begin{split}
H_{P}^{(0)}&=\sum_{i=1}^{2}\Delta\sigma_{n_{i}}^{+}\sigma_{n_{i}}^{-}+\Delta\eta_{\Delta}^{\dag}\eta_{\Delta}\\
&+g_{n_{1}}\sqrt{\frac{2}{L+1}}\left[(\sin{\frac{\pi i}{2}}e^{-i\phi_{n_{1}}}+\cos{\frac{\pi i}{2}}e^{i\phi_{n_{1}}})\sigma_{n_{1}}^{+}\eta_{\Delta}+(\sin{\frac{\pi i}{2}}e^{i\phi_{n_{1}}}+\cos{\frac{\pi i}{2}}e^{-i\phi_{n_{1}}})\eta_{\Delta}^{\dag}\sigma_{n_{1}}^{-}\right],\\
&+g_{n_{2}}\sqrt{\frac{2}{L+1}}\left[(\sin{\frac{\pi j}{2}}e^{-i\phi_{n_{2}}}+\cos{\frac{\pi j}{2}}e^{i\phi_{n_{2}}})\sigma_{n_{2}}^{+}\eta_{\Delta}+(\sin{\frac{\pi j}{2}}e^{i\phi_{n_{2}}}+\cos{\frac{\pi j}{2}}e^{-i\phi_{n_{2}}})\eta_{\Delta}^{\dag}\sigma_{n_{2}}^{-}\right],\\
H_{Q}^{(0)}&=\sum_{k=1,\ne\frac{L+1}{2}}^{L}\epsilon_{c,k}\eta_{c,k}^{\dag}\eta_{c,k}.\\
\end{split}
\end{equation}
The coupling between the two subspaces $P$ and $Q$ is
\begin{equation}
\begin{split}
H_{g}&=\sqrt{\frac{2}{L+1}}\sum_{k=1,\ne\frac{L+1}{2}}^{L}\sin{\frac{\pi k i}{L+1}}(g_{n_{1}}e^{-i\phi_{n_{1}}}\sigma_{n_{1}}^{+}\eta_{c,k}+g_{n_{1}}e^{i\phi_{n_{1}}}\eta_{c,k}^{\dag}\sigma_{n_{1}}^{-})\\
&+\sqrt{\frac{2}{L+1}}\sum_{k=1,\ne\frac{L+1}{2}}^{L}\sin{\frac{\pi k (i+1)}{L+1}}(g_{n_{1}}e^{i\phi_{n_{1}}}\sigma_{n_{1}}^{+}\eta_{c,k}+g_{n_{1}}e^{-i\phi_{n_{1}}}\eta_{c,k}^{\dag}\sigma_{n_{1}}^{-})\\
&+\sqrt{\frac{2}{L+1}}\sum_{k=1,\ne\frac{L+1}{2}}^{L}\sin{\frac{\pi k j}{L+1}}(g_{n_{2}}e^{-i\phi_{n_{2}}}\sigma_{n_{2}}^{+}\eta_{c,k}+g_{n_{2}}e^{i\phi_{n_{2}}}\eta_{c,k}^{\dag}\sigma_{n_{2}}^{-})\\
&+\sqrt{\frac{2}{L+1}}\sum_{k=1,\ne\frac{L+1}{2}}^{L}\sin{\frac{\pi k (j+1)}{L+1}}(g_{n_{2}}e^{i\phi_{n_{2}}}\sigma_{n_{2}}^{+}\eta_{c,k}+g_{n_{2}}e^{-i\phi_{n_{2}}}\eta_{c,k}^{\dag}\sigma_{n_{2}}^{-}).\\
\end{split}
\end{equation}
Since the perturbation Hamiltonian $H_{g}$ (atom-cavity coupling) is off-diagonal, the first order correction $H_{P}^{(1)}=PH_{g}P$ is zero. The second order correction is
\begin{equation}
\begin{split}
H_{P}^{(2)}=-PH_{g}Q\frac{1}{H_{0}-E_{P}^{(0)}}QH_{g}P,\\
\end{split}
\end{equation}
where $E_{P}^{(0)}=\Delta$ is the degenerate energy in $P$ subspace.

The matrix element for $H_{P}^{(2)}$ is
\begin{equation}
\begin{split}
&\langle{n_{\alpha}}|H_{P}^{(2)}|{n_{\beta}}\rangle=-\sum_{k=1,\ne\frac{L+1}{2}}^{L}\frac{\langle{0}|\sigma_{n_{\alpha}}^{-}|H_{g}|\eta_{c,k}^{\dag}|{0}\rangle\langle{0}|\eta_{c,k}|H_{g}|\sigma_{n_{\beta}}^{+}|{0}\rangle}{\epsilon_{c,k}-\Delta},\\
&\langle{\eta_{\Delta}}|H_{P}^{(2)}|{\eta_{\Delta}}\rangle=-\sum_{k=1,\ne\frac{L+1}{2}}^{L}\frac{\langle{0}|\eta_{\Delta}|H_{g}|\eta_{c,k}^{\dag}|{0}\rangle\langle{0}|\eta_{c,k}|H_{g}|\eta_{\Delta}^{\dag}|{0}\rangle}{\epsilon_{c,k}-\Delta},\\
&\langle{n_{\alpha}}|H_{P}^{(2)}|{\eta_{\Delta}}\rangle=-\sum_{k=1,\ne\frac{L+1}{2}}^{L}\frac{\langle{0}|\sigma_{n_{\alpha}}^{-}|H_{g}|\eta_{c,k}^{\dag}|{0}\rangle\langle{0}|\eta_{c,k}|H_{g}|\eta_{\Delta}^{\dag}|{0}\rangle}{\epsilon_{c,k}-\Delta},\\
\end{split}
\end{equation}
where $\alpha,\beta=1,2$. It is straightforward to obtain that
\begin{equation}
\begin{split}
&\langle{0}|\sigma_{n_{\alpha}}^{-}|H_{g}|\eta_{c,k}^{\dag}|{0}\rangle=\\
&\sqrt{\frac{2}{L+1}}\left[g_{n_{1}}(e^{-i\phi_{n_{1}}}\sin{\frac{\pi k i}{L+1}}+e^{i\phi_{n_{1}}}\sin{\frac{\pi k (i+1)}{L+1}})\delta_{1,\alpha}+g_{n_{2}}(e^{-i\phi_{n_{2}}}\sin{\frac{\pi k j}{L+1}}+e^{i\phi_{n_{2}}}\sin{\frac{\pi k (j+1)}{L+1}})\delta_{2,\alpha}\right],\\
&\langle{0}|\eta_{\Delta}|H_{g}|\eta_{c,k}^{\dag}|{0}\rangle=0.\\
\end{split}
\end{equation}
Therefore, the matrix elements for $H_{P}^{(2)}$ are
\begin{equation}
\label{odd_2ndcorr}
\begin{split}
&\langle{n_{1}}|H_{P}^{(2)}|{n_{1}}\rangle=-\frac{g_{n_{1}}^2}{J(L+1)}\sum_{k=1,\ne\frac{L+1}{2}}^{L}\frac{(e^{-i\phi_{n_{1}}}\sin{\frac{\pi k i}{L+1}}+e^{i\phi_{n_{1}}}\sin{\frac{\pi k (i+1)}{L+1}})(e^{i\phi_{n_{1}}}\sin{\frac{\pi k i}{L+1}}+e^{-i\phi_{n_{1}}}\sin{\frac{\pi k (i+1)}{L+1}})}{\cos\frac{\pi k}{L+1}},\\
&=-\frac{g_{n_{1}}^2}{J}((1-(-1)^{i})\frac{L-i}{L+1}+(1+(-1)^{i})\frac{i}{L+1})\cos{2\phi_{n_{1}}},\\
&\langle{n_{2}}|H_{P}^{(2)}|{n_{2}}\rangle=-\frac{g_{n_{2}}^2}{J(L+1)}\sum_{k=1,\ne\frac{L+1}{2}}^{L}\frac{(e^{-i\phi_{n_{2}}}\sin{\frac{\pi k j}{L+1}}+e^{i\phi_{n_{2}}}\sin{\frac{\pi k (j+1)}{L+1}})(e^{i\phi_{n_{2}}}\sin{\frac{\pi k j}{L+1}}+e^{-i\phi_{n_{2}}}\sin{\frac{\pi k (j+1)}{L+1}})}{\cos\frac{\pi k}{L+1}},\\
&=-\frac{g_{n_{2}}^2}{J}((1-(-1)^{j})\frac{L-j}{L+1}+(1+(-1)^{j})\frac{j}{L+1})\cos{2\phi_{n_{2}}},\\
&\langle{n_{1}}|H_{P}^{(2)}|{n_{2}}\rangle=-\frac{g_{n_{1}}g_{n_{2}}}{J(L+1)}\sum_{k=1,\ne\frac{L+1}{2}}^{L}\frac{(e^{-i\phi_{n_{1}}}\sin{\frac{\pi k i}{L+1}}+e^{i\phi_{n_{1}}}\sin{\frac{\pi k (i+1)}{L+1}})(e^{i\phi_{n_{2}}}\sin{\frac{\pi k j}{L+1}}+e^{-i\phi_{n_{2}}}\sin{\frac{\pi k (j+1)}{L+1}})}{\cos\frac{\pi k}{L+1}},\\
&=\frac{g_{n_{1}}g_{n_{2}}}{J}(\cos{\frac{(i+j)\pi}{2}}+\sin{\frac{(i-j)\pi}{2}})[\frac{L-2[\frac{j}{2}]+(-1)^{j}}{L+1}e^{i[(-1)^{i}\phi_{n_{1}}+(-1)^{j}\phi_{n_{2}}]}+\frac{2([\frac{i-1}{2}]+1)(-1)^{i+j-1}}{L+1}e^{-i[(-1)^{i}\phi_{n_{1}}+(-1)^{j}\phi_{n_{2}}]}],\\
&\langle{n_{2}}|H_{P}^{(2)}|{n_{1}}\rangle=-\frac{g_{n_{1}}g_{n_{2}}}{J(L+1)}\sum_{k=1,\ne\frac{L+1}{2}}^{L}\frac{(e^{-i\phi_{n_{2}}}\sin{\frac{\pi k j}{L+1}}+e^{i\phi_{n_{2}}}\sin{\frac{\pi k (j+1)}{L+1}})(e^{i\phi_{n_{1}}}\sin{\frac{\pi k i}{L+1}}+e^{-i\phi_{n_{1}}}\sin{\frac{\pi k (i+1)}{L+1}})}{\cos\frac{\pi k}{L+1}},\\
&=\frac{g_{n_{1}}g_{n_{2}}}{J}(\cos{\frac{(i+j)\pi}{2}}+\sin{\frac{(i-j)\pi}{2}})[\frac{L-2[\frac{j}{2}]+(-1)^{j}}{L+1}e^{-i[(-1)^{i}\phi_{n_{1}}+(-1)^{j}\phi_{n_{2}}]}+\frac{2([\frac{i-1}{2}]+1)(-1)^{i+j-1}}{L+1}e^{i[(-1)^{i}\phi_{n_{1}}+(-1)^{j}\phi_{n_{2}}]}],\\
&\langle{\eta_{\Delta}}|H_{P}^{(2)}|{\eta_{\Delta}}\rangle=0,\\
&\langle{n_{\alpha}}|H_{P}^{(2)}|{\eta_{\Delta}}\rangle=0,\\
\end{split}
\end{equation}
where $i < j$ and $[x]$ denotes the greatest integer less than or equal to $x$. Therefore, the effective Hamiltonian for the odd-sized spin cavity is $H_{\text{eff}} = H_{P}^{(0)} + H_{P}^{(2)}$. We observe that due to the degeneracy between the eigenmodes of the atoms and the $\eta_{\Delta}$ mode in the cavity, the influence of the $\eta_{\Delta}$ mode on the atomic system appears at the first order of the couplings $g_{n_i}$, while the influence of other cavity modes occurs at the second order, $g_{n_i}^2$. This resonance between the cavity mode $\eta_{\Delta}$ and the atomic system results in low concurrence between the atoms in the odd-sized spin cavity.

\subsection{Comparison between the perturbation theory and the numerical simulation}

We will compare the results from the effective Hamiltonian, \( H_{\text{eff}} = H_{P}^{(0)} + H_{P}^{(2)} \), in both even- (Eq. (\ref{2ndcorr})) and odd-sized cavities (Eq. (\ref{odd_2ndcorr})) with numerical simulations based on the full Hamiltonian in Eqs. (\ref{cavity})-(\ref{couplings}).

For the even-sized cavity, the concurrence can be obtained analytically from the effective Hamiltonian. In this case, the concurrence is given by Eq. (\ref{concurr1}) or Eq. (\ref{concurr2}), and for \(\phi = \pi/4\), both expressions reduce to \(C(t) = |\sin{\frac{2g^{2}}{J}t}|\). On the other hand, for the odd-sized cavity, an analytical expression for the concurrence is not straightforward to derive from the effective Hamiltonian, so we use numerical methods to calculate it. For a general state, \(\psi = a|1_{n_{1}}0_{n_{2}}0_{\eta_{\Delta}}\rangle + b|0_{n_{1}}1_{n_{2}}0_{\eta_{\Delta}}\rangle + c|0_{n_{1}}0_{n_{2}}1_{\eta_{\Delta}}\rangle\), the concurrence between the two atoms is given by \(C = 2|ab|\).

\subsubsection{Dynamics of concurrence for both even- and odd-sized cavities}

\begin{figure}[htp]
\includegraphics[width=16cm]{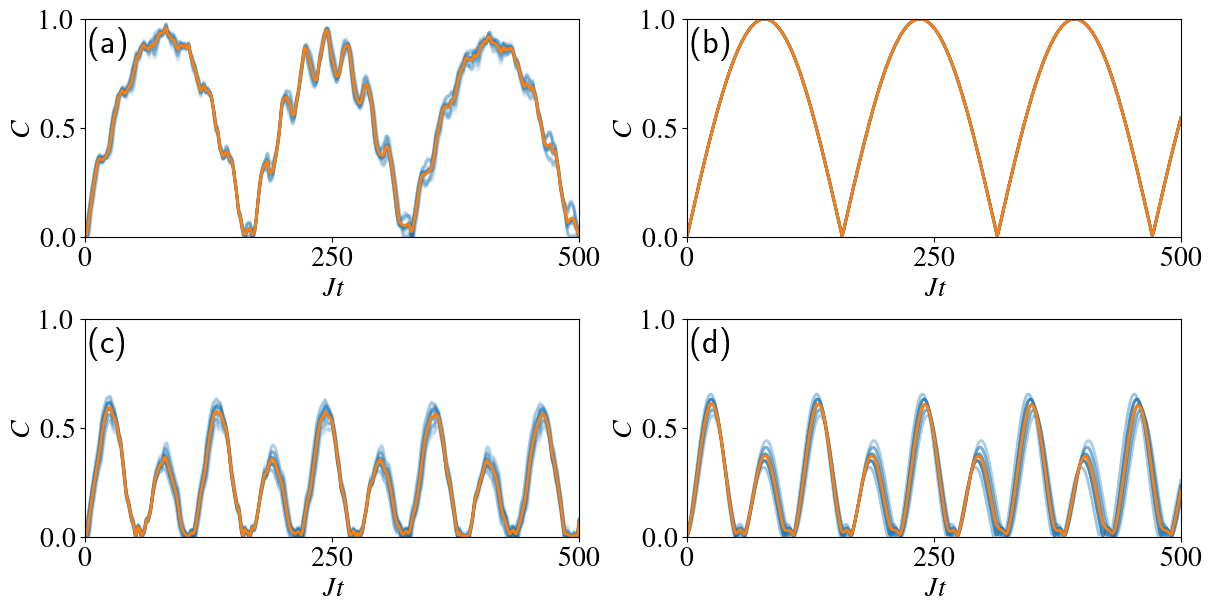}\\
\caption{Dynamics of concurrence as a function of the distance between atoms $\Delta{n}=|\mathcal{L}[n_{2}] - \mathcal{L}[n_{1}]|$. The background blue lines represent the concurrence for all distances between different pairs of atoms, while the orange line shows the average concurrence across these distances. Panels (a) and (b) present the results for the full and effective Hamiltonians for an even-sized cavity with length $L=10$. Panels (c) and (d) present the corresponding results for an odd-sized cavity with length $L=11$.}
\label{comparison}
\end{figure}

Figure \ref{comparison} shows the dynamics of concurrence for the hopping phase \(\phi = \pi/4\) in both even- and odd-sized cavities, comparing results from both the effective and full Hamiltonians. Upon comparison, we find that the perturbation theory within the single-excitation subspace effectively captures the essential physics of the system, with no significant differences observed between the results from the effective and full Hamiltonians.

Additionally, the blue lines in the figure represent the concurrence for various distances between different pairs of atoms, while the orange line shows the average concurrence across these distances. Our findings indicate that for small size cavities, the concurrence is not significantly affected by the distance between the two atoms, particularly in the even-sized cavity. This is clearly seen in Eq. (\ref{concurr1}), where the concurrence derived from the effective Hamiltonian is independent of atom distance. However, when using the full Hamiltonian approach, the concurrence does exhibit some dependence on the distance between the atoms.

\subsubsection{Effects of cavity length, coupling strength and hopping phase to the concurrence}
\label{concorrencebehaviour}

\begin{figure}[htp]
\includegraphics[width=16cm]{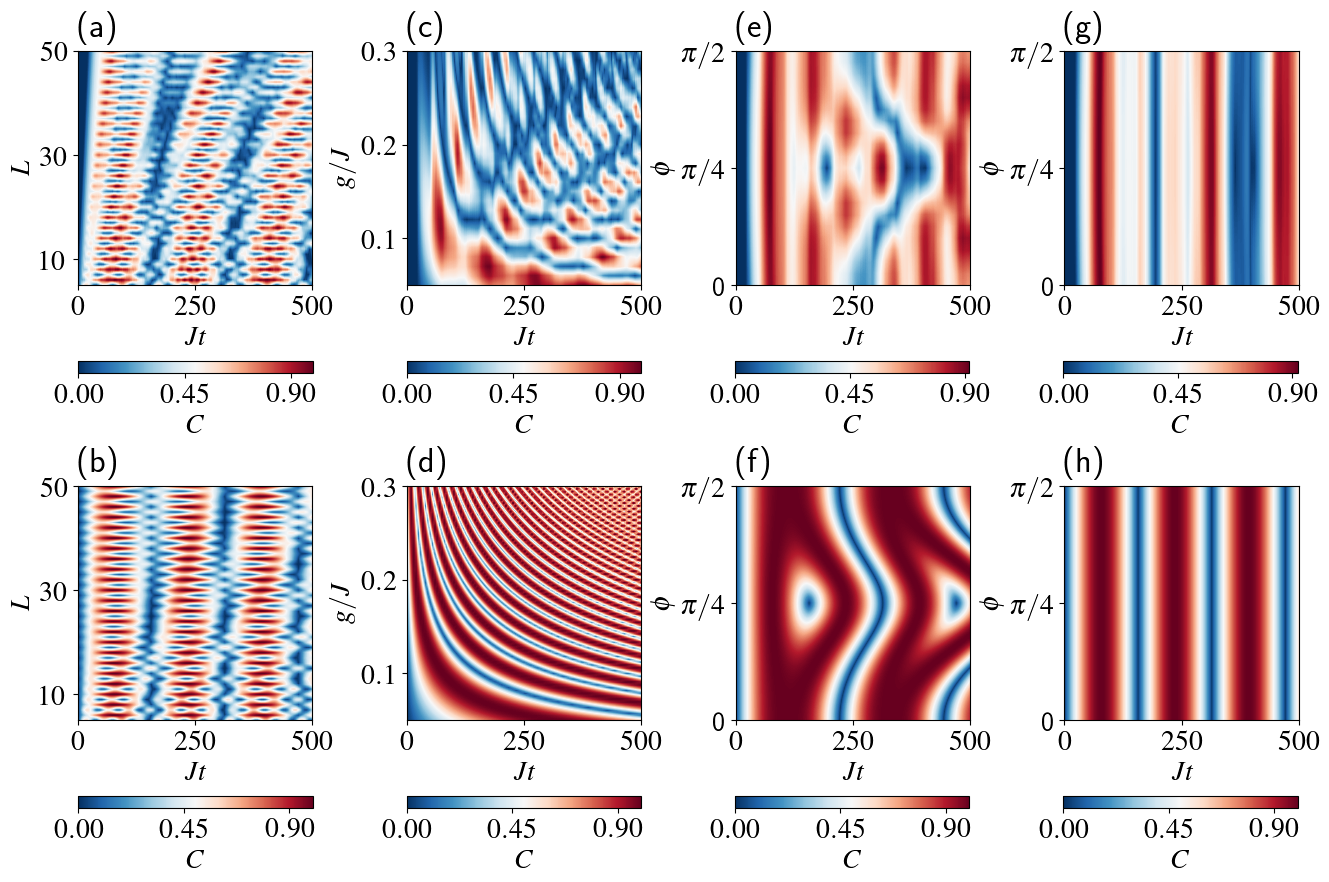}\\
\caption{Effects of cavity length, coupling strength, and hopping phase on the concurrence of two atoms within a spin cavity, comparing the full Hamiltonian approach ((a), (c), (e), (g)) with the effective Hamiltonian approach ((b), (d), (f), (h)). The cavity length is denoted by $L=50$, with two atoms $N=2$ in the cavity. The on-site energy $\Delta$ and driving strength $\Omega_{n_{i}}$ are set to zero. Initially, atom $n_{1}$ is prepared in the excited state $|1\rangle$, and atom $n_{2}$ is in the ground state $|0\rangle$. The hopping amplitude in the spin cavity is $J_{c_{i}}=J=1$, and the coupling between the atoms and the cavity is $g_{n_{i}}e^{-i\phi_{n_{i}}}{\equiv}ge^{-i\phi}$, where $g/J=0.1$.
Panels (a) and (b) explore the effect of cavity length, with $L$ ranging from 5 to 50. The two atoms are positioned at $\mathcal{L}[n_{1}]=2$ and $\mathcal{L}[n_{2}]=L-2$. Panels (c) and (d) examine the effect of coupling strength, where $g/J$ varies from 0 to 0.3. The atoms are placed at $\mathcal{L}[n_{1}]=2$ and $\mathcal{L}[n_{2}]=L-2$. Panels (e) and (f) investigate the impact of hopping phase, with $\phi$ ranging from 0 to $\pi/2$, with the atoms positioned at an odd distance from each other, specifically at $\mathcal{L}[n_{1}]=2$ and $\mathcal{L}[n_{2}]=L-3$. Panels (g) and (h) also study the effect of hopping phase, with the same phase range, but with the atoms positioned at an even distance from each other, specifically at $\mathcal{L}[n_{1}]=2$ and $\mathcal{L}[n_{2}]=L-2$. A bond dimension $D=10$ is used for the MPS simulation.
}
\label{dependence}
\end{figure}

We investigate the effects of cavity length, coupling strength, and hopping phase on the concurrence of two atoms, comparing results obtained from both the full and effective Hamiltonian approaches. Our findings indicate that the two approaches are largely consistent; however, it is important to note that, according to the cavity dispersion \(\epsilon_{c,k}=\Delta+2J\cos\frac{\pi k}{L+1}\), as the cavity size increases, more cavity modes come closer to the atomic modes at energy \(\Delta\). This leads to increased leakage of excitation from the atomic system into these cavity modes, resulting in a decrease in maximum concurrence for larger spin cavities and causing some deviation between the perturbative approach and numerical simulation.

In our study, we consider a spin cavity of length \(L\) with two atoms, as depicted in Fig. \ref{model}. The atoms are located at positions \(\mathcal{L}[n_{1}]\) and \(\mathcal{L}[n_{2}]\) and are initially prepared with atom \(n_{1}\) in the excited state \(|1\rangle\) and atom \(n_{2}\) in the ground state \(|0\rangle\). The on-site energy \(\Delta\) and driving strength \(\Omega_{n_{i}}\) are both set to zero.

Fig. \ref{dependence}(a) and \ref{dependence}(b) examine the effect of cavity length using both the full and effective Hamiltonian approaches, respectively. The cavity length is \(L=50\), with the hopping phase \(\phi_{n_{1}}=\phi_{n_{2}}=\pi/4\). The two atoms are located at positions \(\mathcal{L}[n_{1}]=2\) and \(\mathcal{L}[n_{2}]=L-2\). We observe a parity effect in the system: for odd-sized cavities, a magnonic mode resonates with the atomic subsystem, resulting in low concurrence, whereas for even-sized cavities, there is no resonant cavity mode, leading to higher concurrence.

Fig. \ref{dependence}(c) and \ref{dependence}(d) explore the effect of coupling strength on concurrence. Here, the hopping phase is \(\phi_{n_{1}}=\phi_{n_{2}}=\pi/4\), with atoms located at \(\mathcal{L}[n_{1}]=2\) and \(\mathcal{L}[n_{2}]=L-2\). According to second-order perturbation theory in Sec. \ref{2ndperturb}, the concurrence is given by \(C(t)=|\sin{\frac{2g^{2}}{J}t}|\), as depicted in Fig. \ref{dependence}(d). As the coupling strength increases, the oscillation frequency of concurrence also increases. Numerical results in Fig. \ref{dependence}(c), where the cavity length is \(L=50\), reveal differences between the full and effective Hamiltonian approaches, attributed to higher-order effects. Fig. \ref{dependence}(c) also demonstrates that with increasing coupling strength, the time to achieve maximum concurrence increases, indicating that a weaker atom-cavity interaction is more beneficial for generating entanglement.
This differs from the effective Hamiltonian approach shown in Fig. \ref{dependence}(d), where a strong atom-cavity interaction shortens the time to reach maximum concurrence and enhances entanglement generation.

We further examine the effect of chirality on concurrence, with the degree of chirality controlled by the hopping phase \(\phi_{n_{1}}=\phi_{n_{2}}=\phi\). For a spin cavity of length \(L=50\) with two atoms separated by an odd distance, positioned at \(\mathcal{L}[n_{1}]=2\) and \(\mathcal{L}[n_{2}]=L-3\), Fig. \ref{dependence}(e) shows that as the phase shifts from \(0\) (symmetric coupling) to \(\pi/4\) (chiral coupling), the time required for the first occurrence of maximum concurrence \(Jt_{m}\) decreases for an odd distance \(\Delta{n}=|\mathcal{L}[n_{2}]-\mathcal{L}[n_{1}]|=45\). The concurrence pattern also exhibits symmetry with respect to \(\phi=\pi/4\). Fig. \ref{dependence}(e) further illustrates that, in a spin cavity of length \(L=50\), chiral coupling (\(\phi=\pi/4\)) leads to higher entanglement in a shorter time. Specifically, \(Jt_{m}\) is 73 for \(\phi=\pi/4\), compared to 75 for \(\phi=0\), representing a \(2.7\%\) faster achievement than with symmetric coupling. For a smaller cavity size of \(L=6\) (as discussed in the main text), chiral coupling significantly accelerates entanglement generation by approximately \(50\%\) compared to symmetric coupling. This finding highlights that small even-sized cavities, when chirally coupled to atoms, can significantly enhance and expedite the generation of entanglement. Fig. \ref{dependence}(f) presents the perturbative result from Eq. (\ref{concurr2}), which is qualitatively consistent with the numerical result shown in Fig. \ref{dependence}(e).

On the other hand, for two atoms separated by an even distance, we observe in Fig. \ref{dependence}(g) that the concurrence remains relatively unaffected by changes in the hopping phase \(\phi\). This is also supported by the perturbative result in Eq. (\ref{concurr1}), where the concurrence is shown in Fig. \ref{dependence}(h). However, it is important to note that as the cavity length increases, the perturbative results gradually deviate from the numerical simulations.

\section{Effect of the classical driving field on entanglement}

For the system with driving strengths \(\Omega_{n_{1}} \neq 0\) and \(\Omega_{n_{2}} = 0\), the restriction to the single-excitation subspace is invalid, requiring calculations in the full Hilbert space, which increases the complexity. However, we can still provide a qualitative explanation for the concurrence dips in Fig. 2(b) of the main text.
We consider the decoupled system with \(g_{n_{1}} = g_{n_{2}} = 0\) and study the energy spectrum of each subsystem. We find that the concurrence dips are due to the resonance between the subsystems.

For the driven atom $n_1$ in Eq. (\ref{atoms}), diagonalization reveals that the eigenenergies are \(\epsilon^{\pm}_{n_{1}} = \frac{\Delta \pm \sqrt{\Delta^2 + 4\Omega_{n_{1}}^{2}}}{2}\). For the undriven atom $n_2$ in Eq. (\ref{atoms}), the eigenenergies are \(0\) and \(\Delta\). For the spin cavity subsystem \(H_{c}\) in Eq. (\ref{cavity}), the energy dispersion is given by \(\epsilon_{c,k} = \Delta + 2J\cos\frac{\pi k}{L+1}\).

In Fig. 2(b) of the main text, we examine the even-sized spin cavity. As previously mentioned, there is no cavity mode resonant with the atomic subsystem, resulting in high concurrence. However, by driving atom 1, it is possible to make its excitation energy resonant with the cavity modes in the spin cavity subsystem, leading to the concurrence dips observed in Fig. 2(b). We find that when \(\epsilon^{+}_{n_{1}} - \epsilon^{-}_{n_{1}} = \epsilon_{c,k}\), specifically when \(\Omega_{n_{1}} = \sqrt{J\cos\frac{\pi k}{L+1}\left(\Delta + J\cos\frac{\pi k}{L+1}\right)}\), the driven atom $n_1$ becomes resonant with the cavity mode, resulting in the concurrence dips in Fig. 2(b). Furthermore, in Fig. 2(b), with the on-site energy set to \(\Delta = 0\), the expression simplifies to \(\Omega_{n_{1}} = \left|J\cos\frac{\pi k}{L+1}\right|\). We plot the value of \(\Omega_{n_{1}}\) in Fig. 2(b) and find that it indicates the position of the concurrence dips.

\section{Fast entanglement generation}

\subsection{On-sites energies and hoppings to achieve maximum concurrence $C_m$ for different stopping times $t_f$}

In this section, we give the detailed on-site energies and hoppings to achieve maximum concurrence $C_{m}$ for different stopping times $t_{f}$. From Tables \ref{D0.1}, \ref{D0.4}, and \ref{D1.0}, we observe that, in general, bringing the two atoms into nearly resonance ($\Delta_{n_{1}}\approx\Delta_{n_{2}}$) is beneficial for the entanglement generation, leading to higher concurrence and faster entanglement dynamics.

\begin{table}[htbp]
\centering
\begin{tabular}{|c|c|c|c|c|c|c|c|c|c|c||c|c|}
\cline{2-13}
\multicolumn{1}{c|}{} & $\Delta_{c_{1}}$ & $\Delta_{c_{2}}$ & $\Delta_{c_{3}}$ & $\Delta_{c_{4}}$ & $\Delta_{c_{5}}$ & $\Delta_{c_{6}}$ & $\Delta_{c_{7}}$ & $\Delta_{c_{8}}$ & $\Delta_{c_{9}}$ & $\Delta_{c_{10}}$ & $\Delta_{n_{1}}$ & $\Delta_{n_{2}}$\\
\hline
\textbf{$Jt_{f}=10$} & -0.100 & 0.031 & 0.100 & -0.100 & 0.100 & 0.100 & -0.100 & 0.100 & 0.069 & -0.100 & 0.100 & 0.100 \\
\textbf{$Jt_{f}=15$} & -0.100 & 0.100 & -0.024 & -0.100 & 0.100 & 0.100 & -0.100 & 0.057 & 0.100 & -0.100 & 0.072 & 0.100 \\
\textbf{$Jt_{f}=20$} & -0.100 & 0.100 & -0.100 & -0.100 & -0.100 & 0.013 & -0.100 & -0.099 & 0.100 & -0.100 & 0.072 & 0.100 \\
\textbf{$Jt_{f}=25$} & -0.100 & 0.100 & -0.100 & -0.100 & -0.100 & -0.036 & -0.100 & -0.100 & 0.100 & -0.100 & 0.071 & 0.100 \\
\textbf{$Jt_{f}=30$} & -0.100 & -0.100 & -0.100 & -0.100 & -0.100 & -0.100 & -0.100 & -0.100 & -0.100 & -0.100 & 0.074 & 0.100 \\
\textbf{$Jt_{f}=40$} & -0.100 & -0.100 & -0.100 & -0.100 & -0.100 & -0.100 & -0.100 & -0.100 & 0.022 & -0.100 & 0.095 & 0.100 \\
\textbf{$Jt_{f}=50$} & -0.100 & 0.094 & -0.100 & -0.100 & -0.100 & -0.100 & -0.100 & -0.100 & -0.031 & -0.100 & 0.100 & 0.100 \\
\textbf{$Jt_{f}=60$} & -0.100 & 0.094 & -0.100 & -0.100 & -0.100 & -0.100 & -0.100 & -0.100 & -0.031 & -0.100 & 0.100 & 0.100 \\
\textbf{$Jt_{f}=70$} & -0.100 & 0.094 & -0.100 & -0.100 & -0.100 & -0.100 & -0.100 & -0.100 & -0.031 & -0.100 & 0.100 & 0.100 \\
\textbf{$Jt_{f}=80$} & -0.100 & 0.094 & -0.100 & -0.100 & -0.100 & -0.100 & -0.100 & -0.100 & -0.031 & -0.100 & 0.100 & 0.100 \\
\textbf{$Jt_{f}=90$} & -0.100 & 0.094 & -0.100 & -0.100 & -0.100 & -0.100 & -0.100 & -0.100 & -0.031 & -0.100 & 0.100 & 0.100 \\
\textbf{$Jt_{f}=100$} & -0.100 & 0.094 & -0.100 & -0.100 & -0.100 & -0.100 & -0.100 & -0.100 & -0.031 & -0.100 & 0.100 & 0.100 \\
\hline
\end{tabular}
\caption{Engineering the on-site energies with restriction $r_{\Delta}=0.1$.}
\label{D0.1}
\end{table}

\begin{table}[htbp]
\centering
\begin{tabular}{|c|c|c|c|c|c|c|c|c|c|c||c|c|}
\cline{2-13}
\multicolumn{1}{c|}{} & $\Delta_{c_{1}}$ & $\Delta_{c_{2}}$ & $\Delta_{c_{3}}$ & $\Delta_{c_{4}}$ & $\Delta_{c_{5}}$ & $\Delta_{c_{6}}$ & $\Delta_{c_{7}}$ & $\Delta_{c_{8}}$ & $\Delta_{c_{9}}$ & $\Delta_{c_{10}}$ & $\Delta_{n_{1}}$ & $\Delta_{n_{2}}$\\
\hline
\textbf{$Jt_{f}=10$} & -0.40 & 0.21 & 0.40 & -0.12 & 0.40 & 0.40 & -0.40 & 0.40 & 0.40 & -0.40 & 0.30 & 0.30 \\
\textbf{$Jt_{f}=15$} & -0.40 & 0.04 & 0.40 & -0.40 & -0.40 & -0.02 & -0.40 & 0.40 & 0.40 & -0.40 & 0.40 & 0.39 \\
\textbf{$Jt_{f}=20$} & -0.40 & -0.32 & 0.40 & -0.40 & -0.40 & 0.34 & -0.40 & 0.40 & 0.40 & -0.40 & 0.39 & 0.38 \\
\textbf{$Jt_{f}=25$} & -0.40 & -0.40 & 0.40 & -0.40 & -0.40 & 0.40 & -0.40 & 0.40 & 0.40 & -0.38 & 0.39 & 0.38 \\
\textbf{$Jt_{f}=30$} & -0.40 & 0.40 & 0.15 & -0.40 & 0.40 & 0.40 & -0.40 & 0.26 & 0.40 & -0.40 & 0.14 & 0.18 \\
\textbf{$Jt_{f}=40$} & -0.20 & -0.05 & 0.30 & -0.39 & -0.09 & 0.30 & -0.30 & 0.40 & 0.40 & -0.34 & 0.39 & 0.38 \\
\textbf{$Jt_{f}=50$} & -0.23 & -0.15 & 0.26 & -0.34 & -0.18 & 0.31 & -0.33 & 0.40 & 0.40 & -0.40 & 0.36 & 0.35 \\
\textbf{$Jt_{f}=60$} & -0.11 & 0.40 & -0.18 & -0.40 & 0.24 & -0.29 & 0.11 & -0.04 & 0.13 & -0.29 & 0.11 & 0.11 \\
\textbf{$Jt_{f}=70$} & 0.40 & -0.00 & -0.27 & 0.03 & -0.32 & 0.29 & -0.12 & -0.15 & 0.32 & -0.00 & 0.11 & 0.12 \\
\textbf{$Jt_{f}=80$} & 0.29 & 0.08 & 0.14 & -0.29 & -0.20 & 0.05 & -0.10 & -0.09 & 0.02 & -0.04 & 0.10 & 0.10 \\
\textbf{$Jt_{f}=90$} & 0.27 & 0.25 & -0.05 & -0.35 & 0.02 & -0.16 & 0.00 & -0.24 & 0.13 & -0.08 & 0.08 & 0.08 \\
\textbf{$Jt_{f}=100$} & 0.20 & 0.26 & -0.06 & -0.40 & 0.08 & -0.21 & 0.03 & -0.23 & 0.09 & -0.01 & 0.07 & 0.08 \\
\hline
\end{tabular}
\caption{Engineering the on-site energies with restriction $r_{\Delta}=0.4$.}
\label{D0.4}
\end{table}

\begin{table}[htbp]
\centering
\begin{tabular}{|c|c|c|c|c|c|c|c|c|c|c||c|c|}
\cline{2-13}
\multicolumn{1}{c|}{} & $\Delta_{c_{1}}$ & $\Delta_{c_{2}}$ & $\Delta_{c_{3}}$ & $\Delta_{c_{4}}$ & $\Delta_{c_{5}}$ & $\Delta_{c_{6}}$ & $\Delta_{c_{7}}$ & $\Delta_{c_{8}}$ & $\Delta_{c_{9}}$ & $\Delta_{c_{10}}$ & $\Delta_{n_{1}}$ & $\Delta_{n_{2}}$\\
\hline
\textbf{$Jt_{f}=10$} & -1.00 & 0.63 & 1.00 & 0.28 & 0.60 & 0.60 & -0.21 & 1.00 & 0.76 & -1.00 & 0.51 & 0.51 \\
\textbf{$Jt_{f}=15$} & -1.00 & 0.77 & 1.00 & 0.85 & 1.00 & 0.68 & -1.00 & 0.50 & 1.00 & -1.00 & 0.26 & 0.26 \\
\textbf{$Jt_{f}=20$} & -1.00 & 0.74 & 1.00 & 0.75 & 0.67 & 0.93 & -1.00 & -0.69 & -1.00 & 1.00 & 0.19 & 0.18 \\
\textbf{$Jt_{f}=25$} & -1.00 & 0.68 & 1.00 & 0.71 & 0.34 & 1.00 & -1.00 & -0.86 & -0.99 & 1.00 & 0.10 & 0.10 \\
\textbf{$Jt_{f}=30$} & -0.85 & 0.68 & 0.97 & 1.00 & 0.33 & 1.00 & -1.00 & -0.88 & -1.00 & 0.96 & 0.11 & 0.10 \\
\textbf{$Jt_{f}=40$} & -0.85 & 0.69 & 0.97 & 1.00 & 0.33 & 1.00 & -1.00 & -0.88 & -1.00 & 0.96 & 0.11 & 0.10 \\
\textbf{$Jt_{f}=50$} & -0.85 & 0.69 & 0.97 & 1.00 & 0.33 & 1.00 & -1.00 & -0.88 & -1.00 & 0.96 & 0.11 & 0.10 \\
\textbf{$Jt_{f}=60$} & -0.33 & 0.32 & -0.30 & 0.26 & 0.11 & -0.19 & -0.42 & -0.11 & 0.29 & -0.04 & 0.11 & 0.11 \\
\textbf{$Jt_{f}=70$} & -0.07 & 0.64 & -0.17 & -0.05 & -0.27 & 0.47 & -0.15 & -0.06 & 0.12 & -0.06 & 0.12 & 0.13 \\
\textbf{$Jt_{f}=80$} & 0.29 & -0.84 & -0.21 & -0.15 & 0.09 & -0.05 & -0.05 & -0.17 & -0.06 & -0.12 & 0.10 & 0.11 \\
\textbf{$Jt_{f}=90$} & 0.74 & -0.75 & -0.08 & -0.02 & -0.03 & -0.22 & -0.23 & -0.09 & -0.10 & -0.10 & 0.08 & 0.09 \\
\textbf{$Jt_{f}=100$} & 0.74 & -0.76 & -0.06 & 0.01 & -0.03 & -0.24 & -0.25 & -0.07 & -0.08 & -0.11 & 0.09 & 0.09 \\
\hline
\end{tabular}
\caption{Engineering the on-site energies with restriction $r_{\Delta}=1.0$.}
\label{D1.0}
\end{table}

\begin{table}[htbp]
\centering
\begin{tabular}{|c|c|c|c|c|c|c|c|c|c||c|c|c|c|}
\cline{2-14}
\multicolumn{1}{c|}{} & $J_{c_{1}}$ & $J_{c_{2}}$ & $J_{c_{3}}$ & $J_{c_{4}}$ & $J_{c_{5}}$ & $J_{c_{6}}$ & $J_{c_{7}}$ & $J_{c_{8}}$ & $J_{c_{9}}$ & $g_{n_{1},L}$ & $g_{n_{1},R}$ & $g_{n_{2},L}$ & $g_{n_{2},R}$ \\
\hline
\textbf{$Jt_{f}=10$} &	0.000	&	0.100	&	0.100	&	0.100	&	0.100	&	0.100	&	0.100	&	0.100	&	 0.000	&	0.011	&	0.093	&	0.100	&	0.100	\\
\textbf{$Jt_{f}=15$} &	0.000	&	0.100	&	0.100	&	0.100	&	0.100	&	0.100	&	0.100	&	0.100	&	 0.000	&	0.011	&	0.070	&	0.100	&	0.100	\\
\textbf{$Jt_{f}=20$} &	0.000	&	0.100	&	0.100	&	0.100	&	0.100	&	0.100	&	0.100	&	0.100	&	 0.000	&	0.011	&	0.062	&	0.100	&	0.100	\\
\textbf{$Jt_{f}=25$} &	0.000	&	0.100	&	0.100	&	0.100	&	0.100	&	0.100	&	0.100	&	0.100	&	 0.000	&	0.011	&	0.066	&	0.100	&	0.100	\\
\textbf{$Jt_{f}=30$} &	0.000	&	0.100	&	0.100	&	0.100	&	0.100	&	0.100	&	0.100	&	0.100	&	 0.000	&	0.016	&	0.100	&	0.100	&	0.100	\\
\textbf{$Jt_{f}=40$} &	0.000	&	0.100	&	0.071	&	0.100	&	0.100	&	0.100	&	0.100	&	0.086	&	 0.000	&	0.038	&	0.100	&	0.100	&	0.078	\\
\textbf{$Jt_{f}=50$} &	0.000	&	0.086	&	0.055	&	0.100	&	0.100	&	0.100	&	0.100	&	0.001	&	 0.000	&	0.043	&	0.073	&	0.097	&	0.001	\\
\textbf{$Jt_{f}=60$} &	0.000	&	0.074	&	0.046	&	0.100	&	0.100	&	0.097	&	0.100	&	0.001	&	 0.001	&	0.045	&	0.053	&	0.077	&	0.001	\\
\textbf{$Jt_{f}=70$} &	0.000	&	0.064	&	0.044	&	0.100	&	0.100	&	0.081	&	0.082	&	0.000	&	 0.001	&	0.036	&	0.047	&	0.063	&	0.000	\\
\textbf{$Jt_{f}=80$} &	0.000	&	0.063	&	0.043	&	0.100	&	0.100	&	0.081	&	0.081	&	0.000	&	 0.000	&	0.037	&	0.046	&	0.061	&	0.000	\\
\textbf{$Jt_{f}=90$} &	0.000	&	0.063	&	0.043	&	0.100	&	0.100	&	0.081	&	0.081	&	0.000	&	 0.000	&	0.036	&	0.047	&	0.061	&	0.000	\\
\textbf{$Jt_{f}=100$} &	0.000	&	0.063	&	0.043	&	0.100	&	0.100	&	0.080	&	0.081	&	0.000	&	 0.000	&	0.036	&	0.047	&	0.061	&	0.000	\\
\hline
\end{tabular}
\caption{Engineering the hopping amplitudes with restriction $r_{J}=0.1$.}
\end{table}

\begin{table}[htbp]
\centering
\begin{tabular}{|c|c|c|c|c|c|c|c|c|c||c|c|c|c|}
\cline{2-14}
\multicolumn{1}{c|}{} & $J_{c_{1}}$ & $J_{c_{2}}$ & $J_{c_{3}}$ & $J_{c_{4}}$ & $J_{c_{5}}$ & $J_{c_{6}}$ & $J_{c_{7}}$ & $J_{c_{8}}$ & $J_{c_{9}}$ & $g_{n_{1},L}$ & $g_{n_{1},R}$ & $g_{n_{2},L}$ & $g_{n_{2},R}$ \\
\hline
\textbf{$Jt_{f}=10$} & 0.00 & 0.03 & 0.20 & 0.20 & 0.20 & 0.20 & 0.20 & 0.20 & 0.00 & 0.20 & 0.20 & 0.20 & 0.20 \\
\textbf{$Jt_{f}=15$} & 0.00 & 0.03 & 0.20 & 0.20 & 0.20 & 0.20 & 0.20 & 0.20 & 0.00 & 0.18 & 0.13 & 0.20 & 0.20 \\
\textbf{$Jt_{f}=20$} & 0.00 & 0.20 & 0.14 & 0.20 & 0.20 & 0.20 & 0.20 & 0.17 & 0.00 & 0.08 & 0.20 & 0.20 & 0.16 \\
\textbf{$Jt_{f}=25$} & 0.00 & 0.17 & 0.11 & 0.20 & 0.20 & 0.20 & 0.20 & 0.00 & 0.00 & 0.09 & 0.15 & 0.19 & 0.00 \\
\textbf{$Jt_{f}=30$} & 0.00 & 0.15 & 0.09 & 0.20 & 0.20 & 0.20 & 0.20 & 0.01 & 0.00 & 0.09 & 0.11 & 0.15 & 0.01 \\
\textbf{$Jt_{f}=40$} & 0.00 & 0.10 & 0.09 & 0.19 & 0.20 & 0.13 & 0.14 & 0.08 & 0.00 & 0.05 & 0.09 & 0.08 & 0.06 \\
\textbf{$Jt_{f}=50$} & 0.00 & 0.08 & 0.08 & 0.16 & 0.15 & 0.10 & 0.11 & 0.08 & 0.00 & 0.03 & 0.08 & 0.05 & 0.06 \\
\textbf{$Jt_{f}=60$} & 0.00 & 0.08 & 0.06 & 0.14 & 0.13 & 0.09 & 0.09 & 0.00 & 0.00 & 0.04 & 0.06 & 0.07 & 0.00 \\
\textbf{$Jt_{f}=70$} & 0.00 & 0.07 & 0.06 & 0.13 & 0.12 & 0.08 & 0.08 & 0.00 & 0.00 & 0.03 & 0.06 & 0.07 & 0.00 \\
\textbf{$Jt_{f}=80$} & 0.00 & 0.06 & 0.05 & 0.12 & 0.10 & 0.07 & 0.08 & 0.00 & 0.00 & 0.03 & 0.05 & 0.06 & 0.00 \\
\textbf{$Jt_{f}=90$} & 0.00 & 0.06 & 0.05 & 0.12 & 0.10 & 0.07 & 0.08 & 0.00 & 0.00 & 0.03 & 0.05 & 0.06 & 0.00 \\
\textbf{$Jt_{f}=100$} & 0.00 & 0.06 & 0.05 & 0.12 & 0.10 & 0.07 & 0.08 & 0.00 & 0.00 & 0.03 & 0.05 & 0.06 & 0.00 \\
\hline
\end{tabular}
\caption{Engineering the hopping amplitudes with restriction $r_{J} = 0.2$.}
\end{table}

\begin{table}[htbp]
\centering
\begin{tabular}{|c|c|c|c|c|c|c|c|c|c||c|c|c|c|}
\cline{2-14}
\multicolumn{1}{c|}{} & $J_{c_{1}}$ & $J_{c_{2}}$ & $J_{c_{3}}$ & $J_{c_{4}}$ & $J_{c_{5}}$ & $J_{c_{6}}$ & $J_{c_{7}}$ & $J_{c_{8}}$ & $J_{c_{9}}$ & $g_{n_{1},L}$ & $g_{n_{1},R}$ & $g_{n_{2},L}$ & $g_{n_{2},R}$ \\
\hline
\textbf{$Jt_{f}=10$} & 0.00 & 0.40 & 0.28 & 0.40 & 0.40 & 0.40 & 0.40 & 0.34 & 0.00 & 0.15 & 0.40 & 0.40 & 0.31 \\
\textbf{$Jt_{f}=15$} & 0.00 & 0.30 & 0.18 & 0.40 & 0.40 & 0.39 & 0.40 & 0.01 & 0.00 & 0.18 & 0.21 & 0.31 & 0.01 \\
\textbf{$Jt_{f}=20$} & 0.00 & 0.04 & 0.37 & 0.32 & 0.20 & 0.38 & 0.31 & 0.11 & 0.00 & 0.13 & 0.19 & 0.22 & 0.09 \\
\textbf{$Jt_{f}=25$} & 0.19 & 0.01 & 0.21 & 0.21 & 0.38 & 0.17 & 0.21 & 0.13 & 0.00 & 0.03 & 0.16 & 0.12 & 0.11 \\
\textbf{$Jt_{f}=30$} & 0.00 & 0.20 & 0.22 & 0.13 & 0.31 & 0.23 & 0.20 & 0.12 & 0.00 & 0.15 & 0.00 & 0.11 & 0.10 \\
\textbf{$Jt_{f}=40$} & 0.00 & 0.09 & 0.10 & 0.17 & 0.22 & 0.11 & 0.14 & 0.09 & 0.00 & 0.02 & 0.10 & 0.07 & 0.07 \\
\textbf{$Jt_{f}=50$} & 0.16 & 0.06 & 0.08 & 0.14 & 0.09 & 0.15 & 0.12 & 0.09 & 0.00 & 0.03 & 0.08 & 0.05 & 0.07 \\
\textbf{$Jt_{f}=60$} & 0.00 & 0.08 & 0.06 & 0.14 & 0.13 & 0.09 & 0.09 & 0.00 & 0.00 & 0.04 & 0.06 & 0.07 & 0.00 \\
\textbf{$Jt_{f}=70$} & 0.00 & 0.07 & 0.06 & 0.13 & 0.11 & 0.08 & 0.08 & 0.00 & 0.00 & 0.04 & 0.05 & 0.07 & 0.00 \\
\textbf{$Jt_{f}=80$} & 0.00 & 0.06 & 0.05 & 0.12 & 0.10 & 0.07 & 0.08 & 0.00 & 0.00 & 0.03 & 0.05 & 0.06 & 0.00 \\
\textbf{$Jt_{f}=90$} & 0.00 & 0.06 & 0.05 & 0.12 & 0.10 & 0.07 & 0.08 & 0.00 & 0.00 & 0.03 & 0.05 & 0.06 & 0.00 \\
\textbf{$Jt_{f}=100$} & 0.00 & 0.06 & 0.05 & 0.12 & 0.10 & 0.07 & 0.07 & 0.00 & 0.00 & 0.03 & 0.05 & 0.06 & 0.00 \\
\hline
\end{tabular}
\caption{Engineering the hopping amplitudes with restriction $r_{J}=0.4$.}
\end{table}

\begin{table}[htbp]
\centering
\begin{tabular}{|c|c|c|c|c|c|c|c|c|c||c|c|c|c|}
\cline{2-14}
\multicolumn{1}{c|}{} & $J_{c_{1}}$ & $J_{c_{2}}$ & $J_{c_{3}}$ & $J_{c_{4}}$ & $J_{c_{5}}$ & $J_{c_{6}}$ & $J_{c_{7}}$ & $J_{c_{8}}$ & $J_{c_{9}}$ & $g_{n_{1},L}$ & $g_{n_{1},R}$ & $g_{n_{2},L}$ & $g_{n_{2},R}$ \\
\hline
$Jt_{f}=10$ & 0.00 & 0.86 & 0.55 & 1.00 & 1.00 & 1.00 & 1.00 & 0.00 & 0.04 & 0.43 & 0.73 & 0.97 & 0.00 \\
$Jt_{f}=15$ & 0.00 & 0.30 & 0.62 & 0.78 & 0.56 & 0.70 & 0.67 & 0.02 & 0.00 & 0.06 & 0.20 & 0.21 & 0.02 \\
$Jt_{f}=20$ & 0.00 & 0.18 & 0.20 & 0.32 & 0.43 & 0.23 & 0.27 & 0.11 & 0.00 & 0.05 & 0.20 & 0.18 & 0.09 \\
$Jt_{f}=25$ & 0.00 & 0.14 & 0.16 & 0.27 & 0.34 & 0.19 & 0.21 & 0.11 & 0.00 & 0.04 & 0.16 & 0.14 & 0.09 \\
$Jt_{f}=30$ & 0.00 & 0.08 & 0.25 & 0.51 & 0.47 & 0.13 & 0.22 & 0.16 & 0.00 & 0.01 & 0.15 & 0.08 & 0.13 \\
$Jt_{f}=40$ & 0.00 & 0.14 & 0.17 & 0.10 & 0.24 & 0.18 & 0.16 & 0.13 & 0.00 & 0.11 & 0.02 & 0.04 & 0.10 \\
$Jt_{f}=50$ & 0.16 & 0.07 & 0.09 & 0.15 & 0.09 & 0.16 & 0.13 & 0.09 & 0.00 & 0.03 & 0.09 & 0.05 & 0.07 \\
$Jt_{f}=60$ & 0.00 & 0.08 & 0.06 & 0.15 & 0.13 & 0.09 & 0.09 & 0.00 & 0.00 & 0.04 & 0.06 & 0.07 & 0.00 \\
$Jt_{f}=70$ & 0.00 & 0.07 & 0.06 & 0.14 & 0.11 & 0.08 & 0.08 & 0.00 & 0.00 & 0.04 & 0.05 & 0.07 & 0.00 \\
$Jt_{f}=80$ & 0.00 & 0.06 & 0.05 & 0.12 & 0.10 & 0.07 & 0.08 & 0.00 & 1.00 & 0.03 & 0.05 & 0.06 & 0.00 \\
$Jt_{f}=90$ & 0.00 & 0.06 & 0.05 & 0.12 & 0.10 & 0.07 & 0.08 & 0.00 & 0.76 & 0.03 & 0.05 & 0.06 & 0.00 \\
$Jt_{f}=100$ & 0.00 & 0.06 & 0.05 & 0.12 & 0.10 & 0.07 & 0.08 & 0.00 & 0.00 & 0.03 & 0.05 & 0.06 & 0.00 \\
\hline
\end{tabular}
\caption{Engineering the hopping amplitudes with restriction $r_{J}=1.0$.}
\end{table}

\subsection{Influence of dissipation on fast entanglement generation}

\begin{figure}[htp]
\includegraphics[width=15cm]{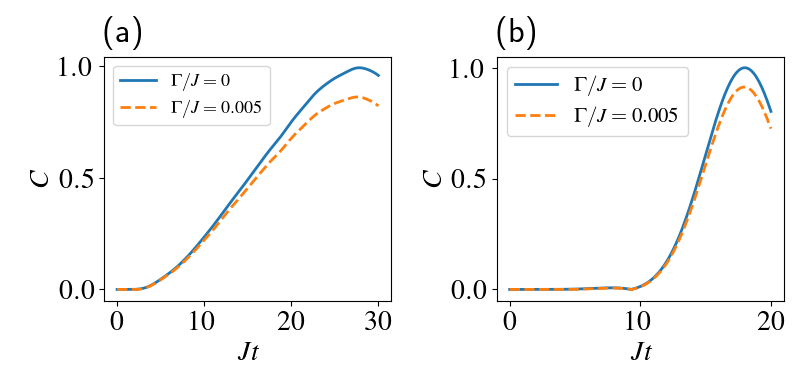}\\
\caption{Fast entanglement generation by engineering the on-site energies or hoppings. The number of cavity spins is $L=10$. The two atoms are located at $\mathcal{L}[n_{1}]=2$ and $\mathcal{L}[n_{2}]=L-2$.
Initially, atom $n_1$ is excited to state $|1\rangle$, while atom $n_2$ is in the ground state $|0\rangle$.
All the cavity spins remain in the ground state.
(a) and (b) Concurrence evolution for two optimal cases with and without dissipation ($\Gamma/J=0.005, 0$). The parameters for (a) are $r_{\Delta}=1.0$, $Jt_{f}=30$, and $\Delta_{c_{i}}/J=$ [$-$0.85, 0.68, 0.97, 1.00, 0.33, 1.00, $-$1.00, $-$0.88, $-$1.00, 0.96] and $\Delta_{n_{i}}/J=$ [0.11, 0.10]. The parameters for (b) are $r_{J}=0.4$, $Jt_{f}=20$, and $J_{c_{i}}/J=$ [0.00, 0.04, 0.37, 0.32, 0.20, 0.38, 0.31, 0.11, 0.00] and $[g_{n_{1},L}, g_{n_{1},R}, g_{n_{2},L}, g_{n_{2},R}]=$ [0.13, 0.19, 0.22, 0.09]. The hopping phases for the two cases are both $\phi_{n_1}=\phi_{n_2}=\pi/4$.
The bond dimension for MPS simulation is $D=30$.}
\label{optimize}
\end{figure}

We consider the influence of dissipation for the two instances in Fig. \ref{optimize}(a) and \ref{optimize}(b), where the dissipation rate $\Gamma_{c_{i}(n_{i})}/J=\Gamma/J=0.005$ is applied to all the sites. We find that the dissipation case (dashed lines) exhibits lower concurrence than the dissipationless case (solid lines). The maximum concurrences reduce to $C_{m}=0.862$ and $C_{m}=0.913$ for engineering of on-site energies and hoppings, respectively.

\subsection{Parameter tolerance for optimizing fast entanglement generation}

In this section, we analyze the parameter tolerance for optimizing fast entanglement generation. Specifically, we investigate deviations in the on-site energies and hopping amplitudes, and we plot the concurrence dynamics to evaluate how these deviations affect the system's behavior compared to the optimal case. Our results indicate that, in most scenarios, the system's performance remains robust against local variations in the parameters, maintaining high concurrence levels despite deviations from the optimal settings.

\begin{figure}[htp]
\includegraphics[width=16cm]{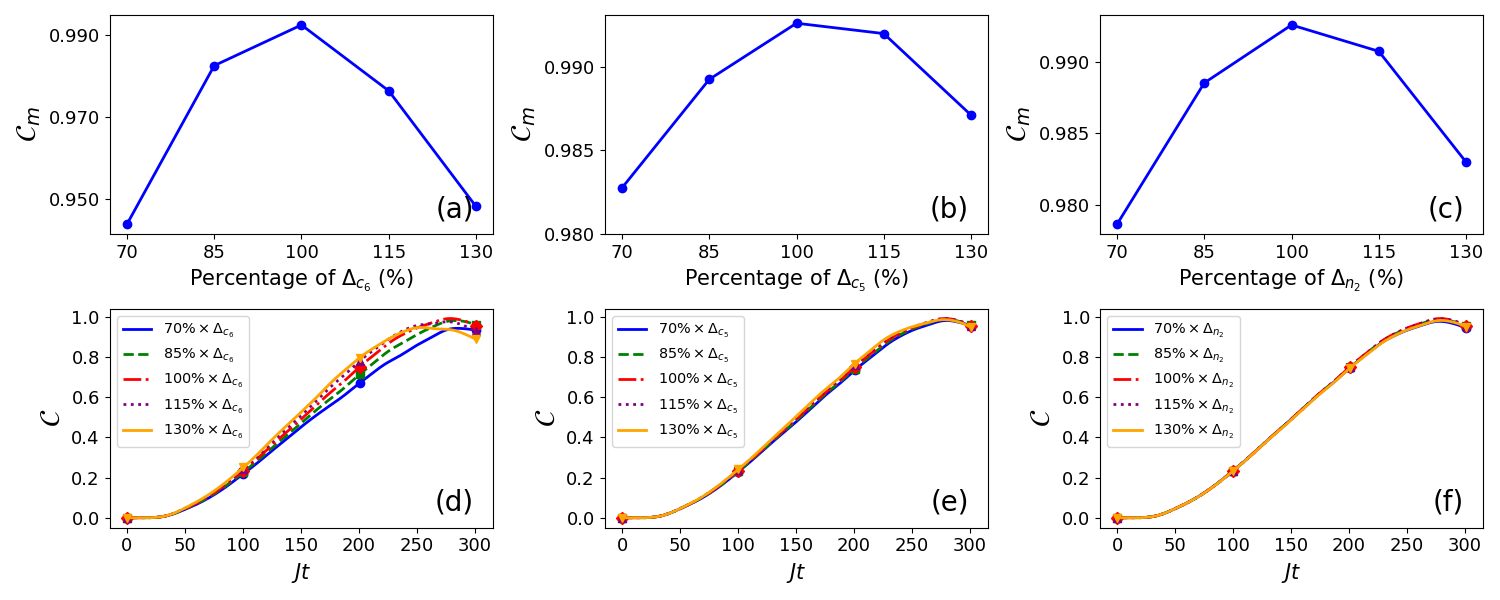}\\
\caption{The impact of on-site energy deviations on maximum concurrence $\mathcal{C}_{m}$ and the concurrence dynamics. Panels (a), (b), and (c) show the maximum concurrence $\mathcal{C}_{m}$ as a function of the percentage deviation of the on-site energies, with deviations ranging from 70$\%$ to 130$\%$. Panels (d), (e), and (f) display the corresponding concurrence dynamics as a function of time for various deviations. The blue lines indicate the results for 70$\%$ of the optimal on-site energy, while the yellow lines represent 130$\%$. The results demonstrate that the system is relatively robust to deviations in the on-site energy, maintaining high concurrence despite changes in on-site energy configuration. The optimal on-site energies are $\Delta_{c_{i}}/J=$ [$-$0.85, 0.68, 0.97, 1.00, 0.33, 1.00, $-$1.00, $-$0.88, $-$1.00, 0.96] and $\Delta_{n_{i}}/J=$ [0.11, 0.10].}
\label{opt_Delta_chiral}
\end{figure}

In Fig. \ref{opt_Delta_chiral}, we explore the effects of varying the on-site energies by up to \(\pm 30\%\). In panels (a), (b), (d), and (e), we examine the cases of maximum and minimum on-site energies of the cavity spins. We present both the maximum concurrence \(\mathcal{C}_{m}\) as a function of the deviation and the corresponding concurrence dynamics over time. The results show that the maximum concurrence is relatively robust to changes in on-site energies, suggesting that small to moderate local deviations from the optimal on-site energy configuration do not significantly impact the entanglement generation.

Additionally, in panels (c) and (f), we study the scenario where the two atoms are tuned progressively closer to resonance. Despite these changes, the optimal concurrence dynamics are not significantly affected, further demonstrating the robustness of the system.

We also analyze the parameter tolerance with respect to deviations in hopping amplitude \(J_{c_{i}}\) and coupling strength \(g_{n_{i}}\), and their impact on fast entanglement generation. By introducing deviations in these parameters, we plot the dynamics of concurrence to observe how closely the system's behavior remains to the optimal case. We find that, similar to the case of on-site energies, the system demonstrates robustness against local variations in both the hopping amplitude and the coupling strength.

\begin{figure}[htp]
\includegraphics[width=16cm]{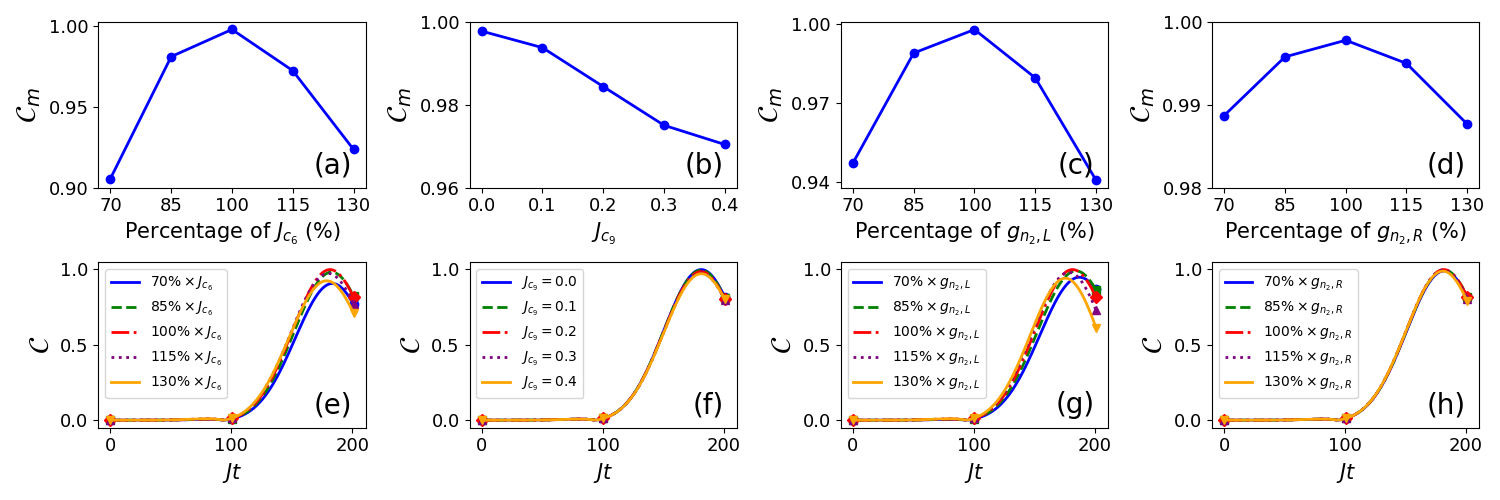}\\
\caption{The impact of deviations in hopping amplitude \(J_{c_{i}}\) and coupling strength \(g_{n_{i},L(R)}\) on maximum concurrence \(\mathcal{C}_{m}\) and the concurrence dynamics. Panels (a), (b), (c), and (d) show the maximum concurrence \(\mathcal{C}_{m}\) as a function of the percentage deviation of \(J_{c_{i}}\), \(g_{n_{i},L}\), and \(g_{n_{i},R}\), with deviations ranging from 70$\%$ to 130$\%$. Panels (e), (f), (g), and (h) display the corresponding concurrence dynamics over time for various deviations. The blue lines represent lower percentages (70$\%$) of the optimal values, while the yellow lines indicate higher percentages (130$\%$). These results demonstrate that the system is robust to moderate deviations in both hopping amplitude and coupling strength, maintaining high concurrence across the parameter variations. The optimal hopping amplitudes and coupling strengths are $J_{c_{i}}/J=$ [0.00, 0.04, 0.37, 0.32, 0.20, 0.38, 0.31, 0.11, 0.00] and $[g_{n_{1},L}, g_{n_{1},R}, g_{n_{2},L}, g_{n_{2},R}]=$ [0.13, 0.19, 0.22, 0.09], respectively.}
\label{opt_J_chiral}
\end{figure}

In Fig. \ref{opt_J_chiral}, we investigate deviations of the hopping amplitude \(J_{c_{i}}\) and coupling strengths \(g_{n_{i},L}\) and \(g_{n_{i},R}\) by up to \(\pm 30\%\). In panels (a), (b), (c), and (d), we show the maximum concurrence \(\mathcal{C}_{m}\) as a function of the percentage deviation in these parameters. The results indicate that while the maximum concurrence decreases slightly as the deviation increases, the system remains largely resilient to parameter variations. Panels (e), (f), (g), and (h) illustrate the concurrence dynamics over time, showing that even with parameter deviations, the system maintains strong entanglement generation, with only minor changes in the concurrence evolution.

These findings suggest that, even with deviations in key parameters, the system retains its ability to generate and sustain high levels of entanglement, making it a promising platform for robust quantum information processing tasks.

\section{Experimental realizations}

In this section, we explore potential experimental realizations of our proposal. We consider two approaches: first, using Rydberg atoms as an analog quantum simulator, and second, utilizing an IBM quantum computer to digitally simulate the time evolution of the Hamiltonian. Each approach offers unique advantages and allows for the practical investigation of the theoretical concepts presented in this work.

\subsection{Analog quantum simulator: Rydberg atom}

\begin{figure}[htp]
\centering
\includegraphics[width=8.5cm]{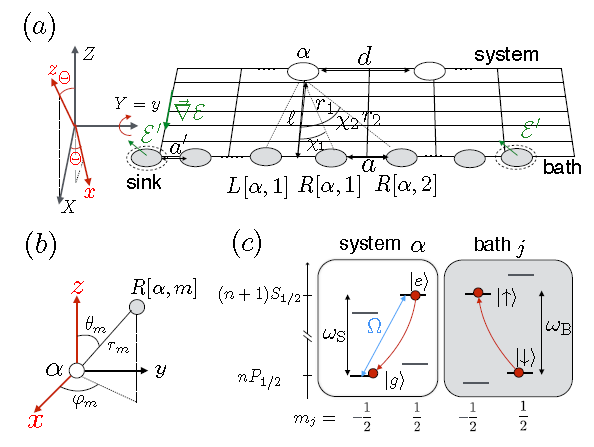}\\
\caption{A Rydberg implementation of the proposed setup.
(a) Rydberg atoms used as the two-level atoms and cavity spins arranged on the $(X,Y)$ plane, where each atom (represented by white disks) engages in dipole-dipole interactions with neighboring cavity spins (represented by gray disks).
(b) The dipole interaction between any two Rydberg atoms, $\alpha$ and $R[\alpha, m]$, is expressed using spherical coordinates $(r_m, \theta_m, \phi_m)$.
(c) This panel shows the energy levels of both atom and cavity spins.
The flip-flop interaction in the triangular plaquette achieved through the dipole-dipole interaction in the Rydberg atoms. (cited from Ref. \cite{PhysRevA.93.063830})
}
\label{Rydberg}
\end{figure}

We present a physical implementation of a chiral network realized through an array of Rydberg atoms as depicted in Fig. \ref{Rydberg} (a). These Rydberg atoms can be arranged using optical lattices, tweezers, or magnetic traps \cite{PhysRevA.93.063830}. To achieve the synthetic gauge field required for chiral coupling, we leverage the inherent ``spin-orbit properties'' of Rydberg dipole-dipole interactions.

In this setup, the hopping phases necessary for chiral coupling are obtained by utilizing the spin-orbit coupling naturally present in the dipole-dipole interactions of Rydberg atoms.
By applying a homogeneous magnetic field in the $z$-direction, we establish a quantization axis and distribute the Rydberg atoms, representing atoms and cavity spins, in the $x$-$y$ plane.
Nearest-neighbor interactions are considered among the Rydberg atoms, with the Hamiltonian for these dipole-dipole interactions expressed through Clebsch-Gordan coefficients and spherical functions.
Specifically, we consider the interactions between an atom and its nearest neighboring cavity spins located immediately to its left and right.

Chirality is achieved by exploiting the spin-orbit coupling contained in the dipole-dipole interactions, resulting in an orbital momentum kick $Y_{2,\mu_{1}+\mu_{2}}(\theta_{m},\varphi_{m}) \propto e^{i(\mu_{1}+\mu_{2})\varphi_{m}}$, where $\mu_{1}+\mu_{2} \neq 0$ and $\varphi_{m}$ is the azimuthal angle shown in Fig. \ref{Rydberg}(b). Atoms and cavity spins are encoded in different magnetic levels $m_{j}$, allowing the transfer of excitation from the atom to the cavity spin associated with the aforementioned momentum kick.
For instance, in Fig. \ref{Rydberg}(c), by choosing specific magnetic levels,
\begin{equation}
\begin{split}
&| e \rangle=|(n+1)S_{1/2},m_{j}=1/2 \rangle , \\
&| g \rangle=|nP_{1/2},m_{j}=-1/2 \rangle , \\
&|\uparrow\rangle=|(n+1)S_{1/2},m_{j}=-1/2 \rangle , \\
&|\downarrow\rangle=|nP_{1/2},m_{j}=1/2 \rangle  . \\
\end{split}
\end{equation}
the flip-flop process $ |e \rangle | \downarrow \rangle \rightarrow |g \rangle | \uparrow \rangle $ leads to a change in angular momentum $\Delta m_{j} = -2$, generating a complex hopping element proportional to $e^{2i\varphi_{m}}$.
Notice that we have neglected the direct dipole-dipole interactions between atoms, which is valid when distance between them is much bigger then the period of cavity spin lattice.
Thus the hopping amplitude between the atom and neighboring cavity spin is given by $\Tilde{J^{m}} = -C_{3}\frac{\sin^{2}{\theta_{m}}}{3r_{m}^{3}}$, where $C_{3}$ is the radial dipole-dipole coefficient, and the hopping phase in our model is $\phi_{m} = 2\varphi_{m}$.

We consider realistic experimental parameters for Rydberg atoms. For Rubidium atoms in the $n = 90$ Rydberg shell, with a dipole-dipole coefficient of $C_{3}= 2\pi \times 65$ $\hbar\text{GHz}\,\mu\text{m}^{3}$, assuming 20 cavity spins separated by 15 $\mu\text{m}$, the nearest-neighbor hopping amplitude is $2\pi \times 2.1$ $\hbar\text{MHz}$. To access the weak coupling regime, setting the distance in $x$-direction between atoms and cavity spins to 34 $\mu\text{m}$ and the magnetic field direction to $\Theta = \pi/3$ yields a weak coupling of $0.07 \times 2\pi \times 2.1$ $\hbar\text{MHz}$, achieving a high degree of chirality with $\gamma_{R} \approx 400\gamma_{L} \approx 2\pi \times 50$ $\text{kHz}$. The long lifetimes of high Rydberg levels, such as $n = 90$, ensure that the decays of atoms can be neglected for practical purposes, with a lifetime of 250 $\mu\text{s}$ compared to $1/\gamma_{R} \approx 3$ $\mu\text{s}$.

\subsection{Digital quantum computer: IBM QX20 Tokyo}

\begin{figure}[htp]
\includegraphics[width=8cm]{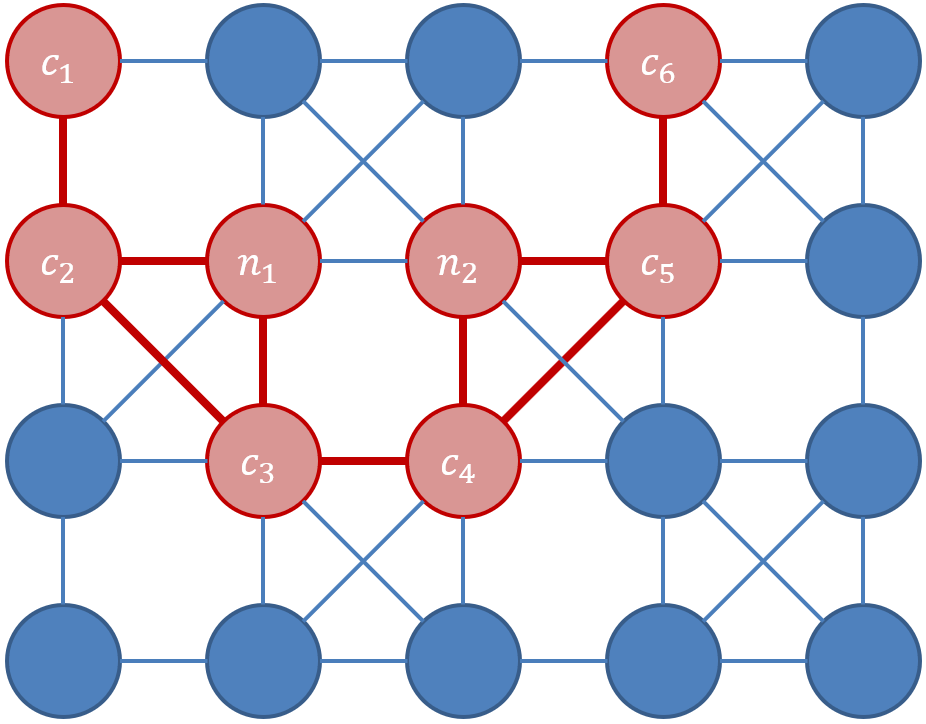}\\
\caption{The connectivity graph of IBM QX20 Tokyo and the subset of qubits to simulate the model.}
\label{QX20}
\end{figure}

We propose using the IBM quantum computer, specifically the IBM QX20 Tokyo \cite{SR.14.13118,Gheorghiu2020ReducingTC,e25030465,10.1007/978-3-030-52482-1_11}, to digitally simulate the time evolution of the model proposed in Fig. \ref{model}.
Fig. \ref{QX20} illustrates the qubit connectivity of IBM QX20 Tokyo, with a focus on a selected subset of qubits highlighted in red. The red lines indicate the connections between these qubits, which are utilized in our simulation. The remaining qubits and connections, shown in blue, represent the broader connectivity of the device, though they are not directly involved in this specific simulation.
This subset of qubits enables us to effectively model the desired Hamiltonian, leveraging the native qubit connections available on the device.

\begin{figure}[htbp]
\includegraphics[width=16cm]{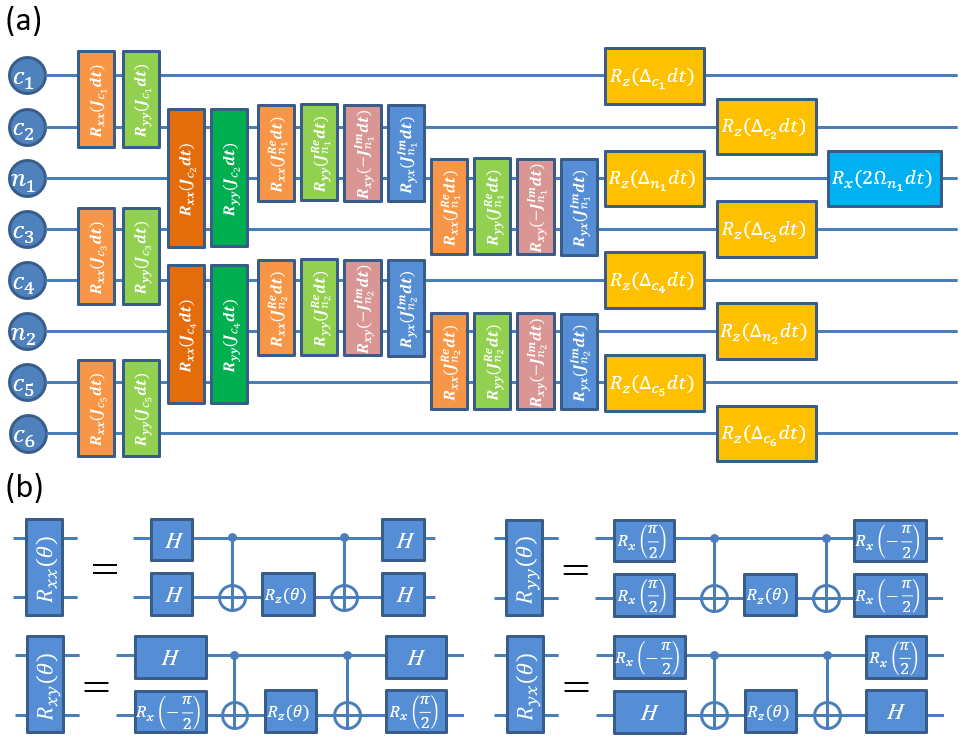}\\
\caption{(a) Quantum circuit for evolving one time step \( dt \) of the Hamiltonian described in Eqs. (\ref{cavity})-(\ref{couplings}). The circuit includes both single-qubit and two-qubit operations in one Trotterization step. (b) Decomposition of the two-qubit gates, \(R_{xx}(\theta)\), \(R_{yy}(\theta)\), \(R_{xy}(\theta)\) and \(R_{yx}(\theta)\), into elementary quantum gates using standard gate sequences.}
\label{circuit}
\end{figure}

The qubits $n_{1}$ and $n_{2}$ are treated as the two atoms, which are coupled to qubits $c_{2}$ and $c_{3}$, and $c_{4}$ and $c_{5}$ as cavity spins, respectively. We then implement the quantum circuit shown in Fig. \ref{circuit} to simulate the model. First, the time evolution operator \(\exp{(-iHT)}\) for Hamiltonian in Eqs. (\ref{cavity})-(\ref{couplings}) is decomposed into single- and two-qubit gates using Trotterization, \((\exp{(-iHdt)})^{T/dt}\). Next, the decomposed gates are sequentially applied to the quantum registers labeled as $c_{1}$, $c_{2}$, $n_{1}$, $c_{3}$, $c_{4}$, $n_{2}$, $c_{5}$, $c_{6}$, completing all necessary operations for one time step $dt$. This process is repeated iteratively until the final time is reached.
Notice that we have chosen the smallest size of system to demonstrate our proposal. Based on the findings in the main text, the small even-sized cavity ($L=6$) optimally enhances the advantages of chirality while also reducing the complexity of the quantum circuit implementation.

\begin{figure}[htbp]
\includegraphics[width=8cm]{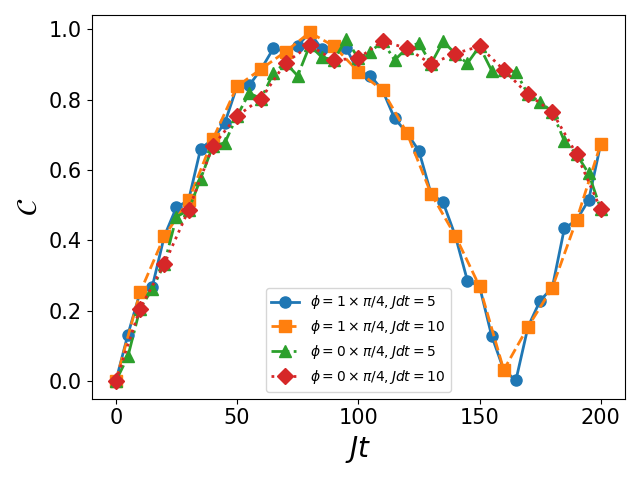}\\
\caption{Dynamics of concurrence \( \mathcal{C} \) as a function of time for different chiralities and time steps. The plot compares the concurrence for two different hopping phase configurations, \( \phi = \pi/4 \) (chiral) and \( \phi = 0 \) (symmetric), and two different time step sizes, \( dt = 5 \) (circle and triangle) and \( dt = 10 \) (square and diamond).}
\label{exp_data}
\end{figure}

We estimate whether the IBM quantum device is sufficient for simulating our model by comparing gate operation times with the coherence times and accounting for the total number of required gate operations. According to Ref. \cite{Nature.618.500} and its Supplementary Information, the device's coherence times are reported to be around 200-300 $\mu\text{s}$ for energy relaxation ($T_{1}$) and approximately 100-200 $\mu\text{s}$ for dephasing ($T_{2}$). The gate operation times are roughly 50 ns for single-qubit gates and around 500 ns for two-qubit gates.

In Fig. \ref{circuit}(a), we present the quantum circuit used to simulate our model, which is composed of two-qubit gates such as \(R_{xx}\), \(R_{yy}\), \(R_{xy}\), \(R_{yx}\), and single-qubit gates like \(R_x\) and \(R_z\). For implementation on quantum devices, Fig. \ref{circuit}(b) shows that each two-qubit gate can be decomposed into three layers of single-qubit gates and two layers of two-qubit gates, which takes approximately \(50\text{ns} \times 3 + 500\text{ns} \times 2 = 1150\text{ns}\). In Fig. \ref{circuit}(a), we observe that each Trotterization step consists of three layers of single-qubit gates and 12 layers of two-qubit gates, which requires about \(50\text{ns} \times 3 + 1150\text{ns} \times 12 \approx 14\mu\text{s}\).

In Fig. \ref{exp_data}, we demonstrate through classical numerical simulation that a time step of \(Jdt=5\), or even \(Jdt=10\), is sufficient to illustrate the advantage of chiral coupling (\(\phi = \pi/4\)) over symmetric coupling (\(\phi = 0\)). This advantage can be observed within \(Jt < 120\), requiring 24 (12) Trotterization steps for \(Jdt=5\) (\(Jdt=10\)), respectively. Therefore, the total gate operation time on the device is approximately \(14\mu\text{s} \times 24 = 336\mu\text{s}\) (\(14\mu\text{s} \times 12 = 168\mu\text{s}\)), both of which fall within the device's coherence time range.

%
%

\clearpage


%

\end{document}